\documentclass[a4paper,11pt]{article}
\pdfoutput=1 
\usepackage{jcappub}

\usepackage[T1]{fontenc} 
\usepackage{placeins}
\usepackage[utf8]{inputenc}
\usepackage{amsmath,mleftright}
\usepackage{xparse}
\usepackage{listings}
\usepackage[dvipsnames]{xcolor}
\usepackage{caption}
\usepackage{subcaption}
\usepackage{graphicx}
\usepackage{dcolumn}
\usepackage{amssymb}
\usepackage{amsmath}
\usepackage{color}
\usepackage{verbatim}
\usepackage{multirow}
\usepackage{soul}
\usepackage{parskip}
\usepackage{mathtools}
\usepackage{enumitem}
\usepackage{tabularx}
\usepackage[normalem]{ulem}  
\usepackage{cleveref}
\graphicspath{
  {./}
  {./figures/}
  {./figures/overlay}
}

\title{Recovering a phase transition signal in simulated LISA data with a modulated galactic foreground}

\subheader{HIP-2024-8/TH}

\author[a,b]{Mark~Hindmarsh}
\author[a]{Deanna~C.~Hooper}
\author[a]{Tiina~Minkkinen}
\author[a]{David~J.~Weir}

\affiliation[a]{Department of Physics and Helsinki Institute of Physics, PL 64, FI-00014 University of Helsinki, Finland}
\affiliation[b]{Department of Physics and Astronomy, University of Sussex, Falmer, Brighton BN1 9QH, U.K}

\emailAdd{mark.hindmarsh@helsinki.fi}
\emailAdd{deanna.hooper@helsinki.fi}
\emailAdd{tiina.minkkinen@helsinki.fi}
\emailAdd{david.weir@helsinki.fi}

\abstract{Stochastic backgrounds of gravitational waves from primordial first-order phase transitions are a key probe of physics beyond the Standard Model. They represent one of the best prospects for observing or constraining new physics with the LISA gravitational wave observatory. However, the large foreground population of galactic binaries in the same frequency range represents a challenge, and will hinder the recovery of a stochastic background. To test the recoverability of a stochastic gravitational wave background, we use the LISA Simulation Suite to generate data incorporating both a stochastic background and an annually modulated foreground modelling the galactic binary population, and the Bayesian analysis code Cobaya to attempt to recover the model parameters. By applying the Deviance Information Criterion to compare models with and without a stochastic background we place bounds on the detectability of gravitational waves from first-order phase transitions. By further comparing models with and without the annual modulation, we show that exploiting the modulation improves the goodness-of-fit and gives a modest improvement to the bounds on detectable models.
}

\begin{document}
\maketitle
\flushbottom

\section{Introduction}
\label{sec:intro}
The creation of catalogues of merging compact binaries by the LIGO, VIRGO, and KAGRA collaborations (LVK)~\cite{KAGRA:2021duu} has solidified gravitational waves as a unique probe of the cosmos. In addition to these compact binaries, there are many cosmological and astrophysical sources that can produce a stochastic gravitational wave background (SGWB, see e.g.~\cite{Christensen:2018iqi,Caprini:2018mtu}). Recent pulsar timing array data from NANOGrav~\cite{NANOGrav:2023gor}, the Chinese Pulsar Timing Array~\cite{Xu:2023wog}, the European Pulsar Timing Array and Indian Pulsar Timing Array~\cite{EPTA:2023fyk} and the Parkes Pulsar Timing Array~\cite{Reardon:2023gzh} (see also Ref.~\cite{InternationalPulsarTimingArray:2023mzf} for a reanalysis of this data) show evidence for a background in the nHz range, 
although as yet there are only upper limits at LVK's peak sensitivity of around 100 Hz~\cite{KAGRA:2021kbb}.
Searching for such a background in the mHz frequency range is one of the science targets of the upcoming Laser Interferometer Space Antenna (LISA) mission~\cite{LISACosmologyWorkingGroup:2022jok}, a space-based GW interferometer set to launch next decade with sensitivity to gravitational waves with frequencies between $10^{-4}$ Hz and $10^{-1}$ Hz~\cite{Audley:2017drz,Colpi:2024xhw}. 

One potential cosmological source of a SGWB in this frequency range is a first order phase transition (PT) in the early universe, first proposed in Refs.~\cite{Witten:1984rs,Hogan:1986dsh}. The phase transition proceeds by the nucleation, expansion, collision and merger of bubbles of the new phase, setting up complex fluid motions which source gravitational waves~\cite{Hindmarsh:2013xza, Hindmarsh:2015qta, Jinno:2016vai, Hindmarsh:2017gnf, Konstandin:2017sat, Cutting:2019zws, Pol:2019yex,Lewicki:2020jiv, Dahl:2021wyk,Jinno:2022mie,Auclair:2022jod,Sharma:2023mao,RoperPol:2023dzg}. 
Gravitational waves are also generated by phase boundary collisions in vacuum phase transitions~\cite{Kosowsky:1991ua,Cutting:2018tjt,Cutting:2020nla,Gould:2021dpm,Lewicki:2020azd,Lewicki:2019gmv}.
The phase transition signal has a characteristically peaked shape which contains information about the parameters of the phase transition such as the latent heat and the supercooling temperature, which in turn are calculable from the underlying physical theory (see~\cite{Hindmarsh:2020hop} for a review). 

The Standard Model has no first order phase transitions, either at the strong interaction  scale~\cite{Borsanyi:2016ksw} or at the electroweak scale~\cite{Kajantie:1996mn}, and therefore does not efficiently source gravitational waves beyond a high-frequency thermal contribution~\cite{Ghiglieri:2020mhm}. Nonetheless, many models beyond the Standard Model predict such a first order PT~\cite{Caprini:2015zlo,Caprini:2019egz}. A possible detection of such a SGWB by LISA would therefore point to new physics beyond the Standard Model, as is required for an explanation of the baryon asymmetry and dark matter of the universe.

There are also astrophysical sources of SGWBs. The strongest in the LISA frequency range is expected to be double white dwarfs in the Milky Way~\cite{Evans:1987qa,Hils:1990vc,Nelemans:2001hp,Karnesis:2021tsh,Boileau:2021sni}, which we refer to henceforth as galactic binaries (GB). Double white dwarfs may also be important extragalactic stochastic source~\cite{Farmer:2003pa,Staelens:2023xjn}, competing with and possibly dominating the expected signal from black hole and neutron star binaries~\cite{Babak:2023lro}.

The detectability of any signal is determined by the instrument noise. In the case of LISA, the instrument is a constellation of three spacecraft exchanging modulated laser signals to determine the relative Doppler shifts of three pairs of freely falling test masses.  The principal noise source is instability in the laser frequencies, which can be greatly reduced by combining the Doppler shifts between different spacecraft at different times, a technique known as Time Delay Interferometry (TDI)~\cite{armstrong1999time,Estabrook:2000ef,Tinto:2003vj,Tinto:2020fcc,Tinto:2022zmf}. There are many other sources of noise, but for modelling purposes they can be collected under two sources each with a characteristic frequency dependence and Gaussian statistics: the test mass motion and the optical metrology system~\cite{Audley:2017drz,LISA_SR_doc}. The test mass acceleration noise has been well characterised by LISA Pathfinder~\cite{Armano:2016bkm}, but other noises must be derived from a detailed physical model of the system~\cite{Bayle:2022okx}.

Here we aim to advance the understanding of the detectability in future LISA data of a PT signal, in the presence of instrument noises and stochastic astrophysical sources in the Milky Way. In particular, we assess how including the annual modulation of the galactic binaries in the signal model~\cite{Adams:2013qma, Boileau:2021sni} improves the ability to separate the signal components, and thereby the sensitivity to a PT signal. 
To make the assessment, we perform Markov Chain Monte Carlo (MCMC) analyses on the simulated datasets, and examine the changes in the Deviance Information Criterion~\cite{Spiegelhalter:2002yvw} when modulation is included in the model. 

We make several simplifications and assumptions, both to reduce the overall computational cost, and to avoid over-parametrising our different signals. We use a two-parameter instrument noise model~\cite{LISA_SR_doc}, a simplified three-parameter GB confusion noise model~\cite{Cornish:2017vip,Wang:2022sti}, and a simplified two-parameter phase transition model~\cite{Hindmarsh:2017gnf,Caprini:2019egz}.  
A phase transition model with a single broken power law is the simplest physical model of a phase transition signal, which however does not account for all the subtleties of source modelling. These additional subtleties could give additional insight into the underlying particle physics, but they would complicate our analysis. Furthermore, by working with the amplitude and frequency of a broken power law signal, we avoid degeneracies between underlying parameters in our statistical analysis.
We allow for annual modulation with a time-dependent GB confusion noise amplitude with two harmonics, which we justify in Appendix~\ref{sec:modulation_parameters}. We do not use more sophisticated methods for anisotropic stochastic sources~\cite{Contaldi:2020rht,Renzini:2018vkx,LISACosmologyWorkingGroup:2022kbp} and we do not attempt to resolve loud sources or perform a global fit (see e.g.~\cite{Littenberg:2020bxy}).  
We also assume that the confusion noise retains its form in the presence of the phase transition signal.

With the exploration of the use of annual modulation, our work extends previous studies of LISA's sensitivity to PTs~\cite{Gowling:2021gcy,Boileau:2021sni,Giese:2021dnw,Boileau:2022ter}. 
Previous work on modulated foregrounds~\cite{Adams:2013qma,Boileau:2021sni} studied the recovery of a cosmological background with a single power-law frequency spectrum, whereas a phase transition signal is expected to be peaked.  We also search specifically for a PT background, rather than a search for a more general cosmological background~\cite{Karnesis:2019mph,Caprini:2019pxz,Flauger:2020qyi}.
Another difference is the use the LISA Simulation Suite~\cite{Bayle:2023qfo}, 
with which we can create simulated time series TDI data featuring instrument noises, confusion noise coming from galactic binaries, and an injected PT signal. 
Using the LISA Simulation Suite offers several advantages: more realistic handling of the instrument noise, better control of the galactic binary noise, and having a systemic and reproducible set-up based on publicly-available codes (but see Refs.~\cite{Banagiri:2021ovv,Rieck:2023pej} for an alternative approach combining detector simulation and Bayesian inference). 

We find that including annual modulation of the amplitude of the GB confusion noise 
does increase the sensitivity to the cosmological stochastic background from a PT, except for those whose peak frequency is within about 20\% of the confusion noise peak frequency. 
In these cases, the PT and GB signals are too similar to be disentangled, regardless of the inclusion of the modulation parameters. 

This paper is organised as follows. We begin by discussing the production and processing of simulated LISA data in Sections~\ref{sec:data} and~\ref{sec:proc}, respectively. In Section~\ref{sec:stats} we derive our likelihood function and review the set-up of our MCMC analyses, before discussing our results in Section~\ref{sec:results} and concluding in Section~\ref{sec:conclusions}. Finally, in Appendix~\ref{sec:pipeline} we present the necessary steps to reproduce our simulated data; while in Appendix~\ref{sec:modulation_parameters} we discuss the choices we made for the Fourier expansion used for the annual modulation; and in Appendix~\ref{sec:dDICvalues} we provide tables with all of the numerical values for our Figs.~\ref{fig:scatter_0vsP} and~\ref{fig:scatter_McvsMPc}.

\section{Simulated LISA data}
\label{sec:data}

To generate a time series of data that resembles what one might expect from LISA, we use the LISA Simulation Suite, which is a Python-based simulation pipeline covering the different elements of the LISA mission~\cite{Bayle:2023qfo}. 
In our simulated datasets, we inject two signals: a stochastic background with broken power-law form resembling what would be produced by sound waves from a first order phase transition, and a foreground resembling the galactic binary foreground population.

The detector response to these signals is computed by the tool \texttt{LISA GW Response}~\cite{LISAGWResponse}, which also takes in the orbital information in the form of an orbit file from \texttt{LISA Orbits}~\cite{LISAOrbits}. Here we use Keplerian orbits. The output is then fed into \texttt{LISA Instrument}~\cite{LISAInstrument}, where we can select the noise sources, and adjust various other properties relating to signal processing on the instrument~\cite{Bayle:2022okx}. The output is a time series of beatnote measurements. In the final step of our simulation, these are given as input to \texttt{PyTDI}~\cite{PyTDI}, which performs the TDI calculations, i.e. the Doppler shift combinations that suppress laser noise. We opt to use second generation TDI as it takes into account the laser arm length fluctuations~\cite{Estabrook:2000ef,Tinto:2002de,Tinto:2003vj}. More specifically, we produce the Michelson $X_2$ combination. 

In the following, we introduce our GW sources and the signals we expect from them. We also elaborate on the detector response and noise functions, writing out the analytical expressions to be used in our data analysis. Lastly, we outline the TDI calculations leading to laser noise cancellation. For a more detailed description of our simulation for data reproduction purposes, refer to Appendix~\ref{sec:pipeline}. Additionally, we have made our full data simulation and analysis pipeline publicly available.\footnote{\url{zenodo.org/doi/10.5281/zenodo.12781278}}

\paragraph{Injected signal: Galactic binaries:}
We start by using the catalogue of galactic white dwarf binaries (GBs) created for the LISA Data Challenge Sangria (LDC2a)~\cite{LDC2a}, which consists of three sets of GBs: verification, interacting, and detached binaries, totalling $\sim 29$ million binaries. The distribution is derived from the population synthesis model by Ref.~\cite{Nelemans:2000es}, and it is concentrated around the bulge and the plane of our galaxy, as illustrated in Fig.~\ref{fig:skymap48}. Note that there are various different binary population models, some of which may have more isotropic or asymmetrical distributions, such as that presented in Ref.~\cite{Lamberts:2019nyk}.

Rather than simulating each of these binaries individually, which would be computationally expensive, we instead create a skymap representing the spatial distribution of binaries in LDC2a. We then insert a function describing the confusion noise coming from the binary population into the pixels of the skymap.
We justify and elaborate on this choice in Appendix~\ref{sec:pipeline}.
There are various analytical functions obtained by fits to population models~\cite{Cornish:2017vip, Karnesis:2021tsh, Wang:2022sti}, where the confusion noise foreground has been shown to peak at around 2 mHz, with estimations for the peak amplitude ranging from $10^{-46}$ to $10^{-44}\; \mathrm{Hz}^{-1}$. 
These confusion noise models are for the unresolved GB, as GBs with high signal-to-noise ratio are expected to be individually resolvable by LISA, and their inclusion would significantly enhance the overall signal~\cite{Littenberg:2023xpl, Katz:2024oqg, Strub:2024kbe}.
In order to reduce the number of parameters while still maintaining a close fit to the unresolved GB data in the frequency range of interest, we choose the model from Ref.~\cite{Wang:2022sti}:
\begin{equation}
  S_\mathrm{gb}(f) = A_\mathrm{gb}f^{\alpha_\mathrm{gb}}[1+\tanh(\gamma_\mathrm{gb} (f_\mathrm{gb}-f))] \, ,
  \label{CPSD}
\end{equation}
where $A_\mathrm{gb}=1.4\times 10^{-45}\; \mathrm{Hz}^{-1}$ is the peak amplitude, $f_\mathrm{gb}=1.29$ mHz is the peak frequency, $\alpha_\mathrm{gb}$ is the spectral slope, and $\gamma_\mathrm{gb}=900$ is a fit parameter that depends on the properties of the GB population. Here we assume that the spectral shape is the same for all binary populations across the sky. As in the aforementioned references, we fix $\alpha_\mathrm{gb} = -7/3$ as per the theoretical prediction for gravitational wave emission from a binary population~\cite{Cornish:2017vip}.

\subparagraph{Annual modulation of galactic binaries:}

\begin{figure}[t]
  \centering
  \includegraphics[width=0.95\textwidth]{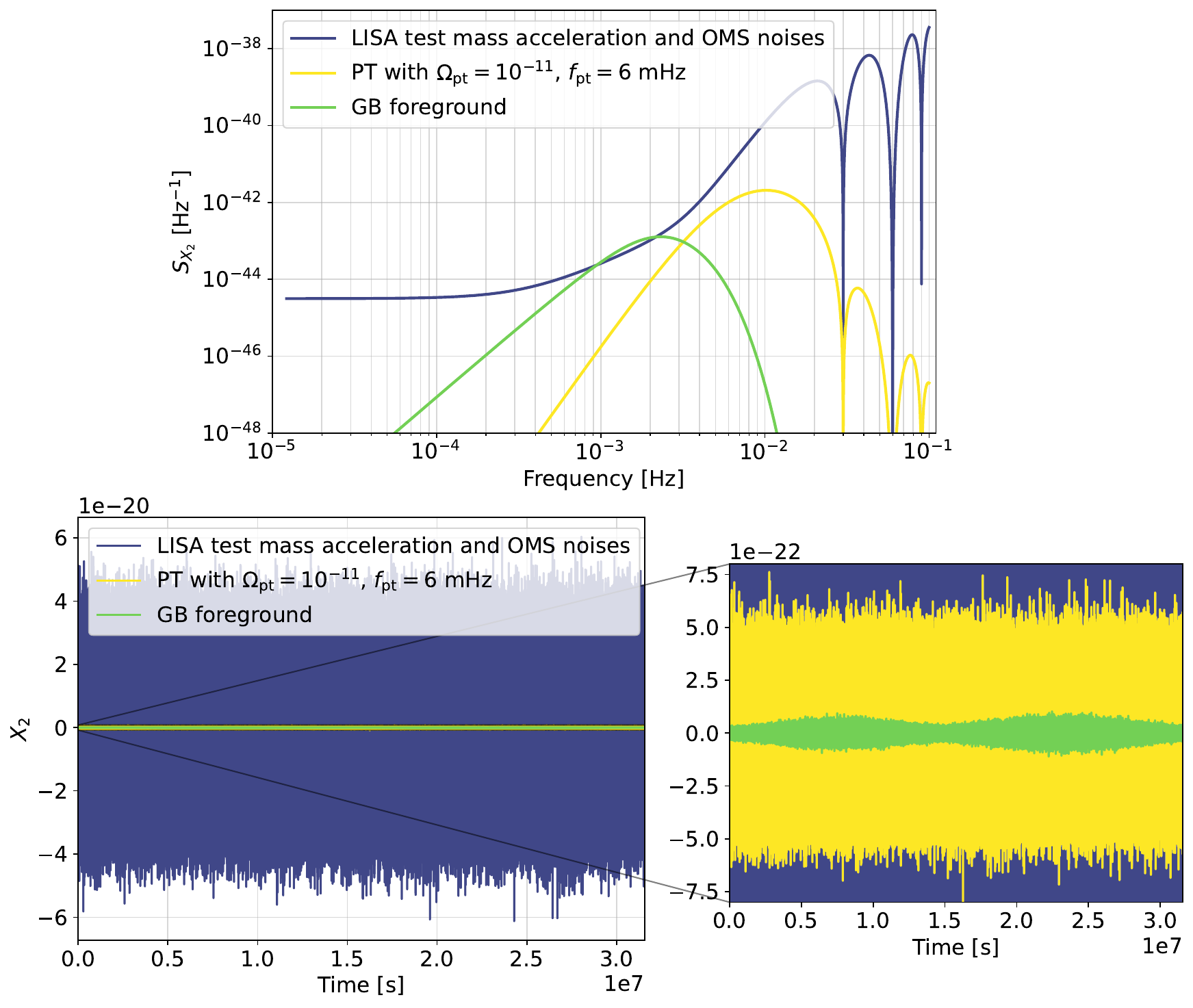}
  \caption{
  Power spectral density (top) and time series of the TDI variable $X_2$ (bottom) of the LISA instrument noise (blue), an example injected PT model with $\Omega_{\mathrm{pt}}=10^{-11}, \ f_{\mathrm{pt}}=6\times 10^{-3}$ Hz (yellow), and the GB confusion noise (green). The bottom right panel shows a zoomed in version of the same time series.
  }
  \label{fig:data}
\end{figure}

LISA's response to gravitational waves is anisotropic: waves incident from directions normal to the plane of the spacecraft constellation induce a larger response than those arriving in the plane.  The plane of the constellation is inclined with respect to the plane of LISA's orbit around the sun, and hence the response to a source with fixed sky position is modulated with period of one year. Galactic binaries are concentrated near the galactic plane, and therefore the amplitude of the GB confusion noise will naturally be modulated, and can be represented in Fourier modes with period $t_\text{yr}$, the sidereal year:
\begin{equation}
  S_\text{gb}(f,t) = S_\mathrm{gb}(f)\left[1 + \sum_{n=0}^{n}  A_n\cos(\omega_n t + \phi_n)\right] \, , 
  \label{eq:gb_fourier}
\end{equation}
where $\omega_n = 2\pi n/t_\text{yr}$ and $A_n$ are the Fourier coefficients (for a full derivation of these, see e.g.~\cite{Giampieri:1997}).
The broad beam of the response function sweeps across the galaxy twice per year in different sky locations, and hence we expect the principal modulation to be in the first and second harmonics, as shown in e.g.~\cite{Giampieri:1997, Digman:2022jmp}. Our simulated GB signal (see Fig.~\ref{fig:data}) clearly shows this effect. We therefore construct our model for the modulated GB confusion noise as 
\begin{equation}
  S_\text{gb}(f,t) = S_\mathrm{gb}(f)\left[1 + A_1\cos(\omega_1 t + \phi_1) + A_2\cos(\omega_2 t + \phi_2)\right] \, .
  \label{eq:mod}
\end{equation}
We investigate the effect of including higher harmonics in Appendix~\ref{sec:modulation_parameters}. We find that their inclusion does not significantly contribute to the GB modulation, and does not improve the ability to resolve a PT signal.

\paragraph{Injected signal: First order phase transition:}
The PT signal is expected to be peaked at wavelengths around the mean bubble spacing (see e.g.~\cite{Hindmarsh:2020hop} for a review), which must be less than the Hubble radius at the time of the phase transition. The amplitude and detailed shape of the PT power spectrum as a function of the physical parameters of the phase transition is still an evolving field (for recent work see~\cite{Sharma:2023mao,RoperPol:2023dzg}).  

Here we use a simplified model for the gravitational wave power spectrum, obtained from fitting to numerical simulations of phase transitions~\cite{Hindmarsh:2017gnf}
\begin{equation}
  \Omega_{\mathrm{gw}}(f) = \Omega_{\mathrm{pt}} \mathcal{P} (f) \, ,
  \label{PTomega2}
\end{equation}
where the spectral shape 
function is 
\begin{equation}
  \mathcal{P}(f) = \left(\frac{f}{f_{\mathrm{pt}}}\right)^3 \left(\frac{7}{4+3(f/f_{\mathrm{pt}})^2} \right)^{7/2}\label{PTPSD}
\end{equation}
and $f_{\rm pt}$ is the peak frequency.  
The GW power spectrum is related to the power spectral density at the detector $S_\text{pt}$ by 
\begin{equation}
\Omega_{\mathrm{gw}}(f) = \frac{4 \pi^2}{3} \frac{f^3}{H_0^2} S_\text{pt}(f).
\end{equation}
where $H_0$ is the Hubble rate at the current epoch.
A phase transition at temperature $T_*$ when the Hubble rate is $H_*$, with mean bubble spacing $R_*$ would have 
\begin{equation} 
{\label{eq:f0} }
f_{\rm pt } \simeq  10^{-6}(H_*R_*)^{-1} \left({T_*}/{100\,\textrm{GeV}}\right)  \,\textrm{Hz},
\end{equation}
The amplitude $\Omega_{\mathrm{pt}}$ depends on $H_*R_*$ and on other parameters of the phase transition (see e.g.~\cite{Caprini:2019egz} for a discussion). Current modelling predicts that phase transitions in the interesting temperature range 100 GeV to 1 TeV can give peak frequencies between $10^{-4}$ Hz and $10^{-2}$ Hz and peak amplitudes in the range $10^{-14}< \Omega_{\mathrm{pt}}< 10^{-9}$~\cite{Gowling:2021gcy}.

While there is a lot of information about the phase transition contained in the detailed shape of the signal which can in principle be recovered~\cite{Gowling:2022pzb}, our concern here is detectability, where the shape is not of primary importance beyond the presence of a peak. 
The $f^3$ behaviour just below the peak is expected for phase transitions with mean bubble spacing of order the Hubble radius at time of the phase transition~\cite{Sharma:2023mao,RoperPol:2023dzg}, which are the loudest signals. The high frequency $f^{-3}$ behaviour characteristic of shocked fluids~\cite{Hindmarsh:2016lnk,Dahl:2021wyk} 
generally emerges only for $f \gg f_\text{pt}$; the $f^{-4}$ behaviour in Eq.~\eqref{PTPSD} is an approximation near the peak to the domed shape seen in numerical simulations~\cite{Hindmarsh:2017gnf}.
We leave the investigation of more complex spectral shapes for future work.

\paragraph{LISA Instrument noise:}
The expected instrument noises for LISA roughly fall into two categories: optical metrology system (OMS) noises and acceleration noises~\cite{Armano:2016bkm}. The former includes shot noise, clock noise, residual laser noise, and beam-pointing instabilities, while the latter comprises different effects accelerating the test masses, such as thermal effects, gravitational forces from surrounding bodies, and electrical forces. The optical-path noises are the dominant component in the higher frequencies, while the acceleration noises dominate in the lower-frequency end of the LISA spectrum. The cut-off frequency is at around $2\times 10^{-3}$ Hz.

The acceleration noise contribution, derived from the LISA Pathfinder results~\cite{Armano:2016bkm}, has a power spectral density
\begin{equation}
  S_{\mathrm{Acc}}(f)=A_{\mathrm{Acc}}^2 \left(\frac{1}{2\pi fc}\right)^2 \left[1+\left(\frac{f_{\mathrm{Acc}}}{f}\right)^2\right] \left[1+\left(\frac{f}{8\times 10^{-3}\;\mathrm{Hz}}\right)^4\right] \, ,\label{Sacc}
\end{equation}
where $A_{\mathrm{Acc}}=3\times 10^{-15}\;\mathrm{m}\,\mathrm{s^{-2}}\, \sqrt{\text{Hz}^{-1}}$, is the amplitude spectral density (ASD), $f_{\mathrm{Acc}}=4\times 10^{-4}\;\mathrm{Hz}$ is the cutoff frequency, and the factor $\left(1/(2\pi fc)\right)^2$ is included to convert the quantity from units of acceleration to relative frequency fluctuations.

The power spectral density for the OMS noise contribution is given by
\begin{equation}
  S_{\mathrm{OMS}}(f)=A_{\mathrm{OMS}}^2 \left(\frac{2\pi f}{c}\right)^2 \left[1+\left(\frac{f_{\mathrm{OMS}}}{f}\right)^4\right]\label{OMS}
\end{equation}
where $A_{\mathrm{OMS}}=15\times 10^{-12}\;\mathrm{m}\, \sqrt{\text{Hz}^{-1}}$, $f_{\mathrm{OMS}}=2\times 10^{-3}\;\mathrm{Hz}$, and again there is an additional factor of $\left(2\pi f/c\right)^2$ to convert from units of displacement to relative frequency fluctuations.

These two dominant noises enter the \texttt{LISA Instrument} simulation in the forms given above~\cite{Bayle:2022okx}, and we use this information when we formulate our full noise model for the analysis of the TDI $X_2$ channel data.

\paragraph{Time-delay interferometry:}
Laser frequency noise is the largest source of noise. In ground-based observatories, where the interference is between split beams from a single laser, the laser frequency noise cancels out. In LISA, the large arm lengths mean that the beams are not sufficiently intense to make a return trip between spacecraft, and so interference is between different lasers. This poses a challenge that can be overcome by time-delay interferometry (TDI)~\cite{Armstrong:2003ut,Tinto:2020fcc}, where interferometer measurements are time-shifted and combined ways that lead to the cancellation of laser noise.

The measurements exchanged between spacecraft can be considered either as phase shifts or as a fractional frequency differences, constructed from interferometer measurements on the pairs of optical benches. A gravitational wave passing between the spacecraft induces a time-dependent change in the path length, and hence a frequency shift. As outlined above, frequency shifts are also introduced by unmodelled acceleration of the test masses, and by the measurements on the optical benches themselves.

Here we consider the fractional frequency shift $\eta_{ij}$ between data arriving on spacecraft $i$ from spacecraft $j$ along the laser arm of length $L_{ij}$. TDI variables are constructed from linear combinations of the signals with delay operators $D_{ij}\eta(t)=\eta_{ij}(t-L_{ij})$ applied. Multiple or nested delays can be represented with the notation $D_{i_1i_2 \ldots i_n} = D_{i_1i_2} D_{i_2i_3} \ldots D_{i_{n-1} i_n}$. Here, we follow the notation of Ref.~\cite{Nam:2022rqg}, without distinguishing between the true arm length and the estimated arm length.

First-generation TDI cancels the laser noise when the arm lengths are constant~\cite{armstrong1999time}.  Three independent noise-cancelling combinations can be taken. A common set is denoted $X$, $Y$ and $Z$, where 
\begin{equation}
X = (1 - D_{131}) (\eta_{12} + D_{12}\eta_{21}) - (1 - D_{121}) (\eta_{13} + D_{13}\eta_{31}),
\end{equation}
with $Y$ and $Z$ obtained by cyclic permutation.
These combinations construct a noise-cancelled Michelson interferometer out of pairs of arms, which may be of unequal length. A common simplifying assumption in modelling is that the arms are of equal length, but searches for stochastic backgrounds in first generation TDI with unequal length arms, as well as different noises in each of the spacecraft, have recently been considered~\cite{Hartwig:2023pft}.

The arm length also changes during the orbit, introducing uncancelled laser noise into the data~\cite{Cornish:2003tz}. The laser noise can again be cancelled with second-generation TDI~\cite{Shaddock:2003dj,Tinto:2022zmf}, which we use here. This involves more round trips of the laser along the arms. The second-generation Michelson $X_2$ variable is
\begin{align}
X_2 &= 
(1 - D_{12131}) \left[ \eta_{13} + D_{13}\eta_{31} + D_{131}(\eta_{12} + D_{12}\eta_{21})\right] \nonumber \\
&-
(1 - D_{13121}) \left[ \eta_{12} + D_{12}\eta_{21} + D_{121}(\eta_{13} + D_{13}\eta_{31})\right]. 
\end{align}
%


\paragraph{Detector Response:}
As outlined above, a simulated time stream $X_2$ containing the detector response to the GW signals from galactic binaries and phase transitions, as well as the instrument noises, is constructed using the tools \texttt{LISA GW Response}, \texttt{LISA Orbit}, \texttt{LISA Instrument} and \texttt{PyTDI}. 
Our data analysis model is formulated in frequency space, and so we need a representation of the detector responses as a function of frequency.  There is no general representation of second-generation TDI response functions in closed form, but closed-form approximations exist in the limit of constant and equal arm lengths. With constant arm lengths, the time delays $D_{ij}$ and $D_{ij}$ commute, and 
\begin{equation}
X_2 \simeq (1 - D_{12131}) X
\label{eq:tdi2approx}
\end{equation}
If the arm lengths are equal, closed-form expressions for the $X$ response functions exists, and the TDI-2 response is then just obtained by multiplying by the modulus squared of the Fourier transform of $1 - D_{12131}$~\cite{Nam:2022rqg}, or $4\sin^2(2 f/f_*)$, where $f_*=c/(2\pi L)$ is the transfer frequency and $L$ is the common arm length.  For the nominal LISA arm length of $2.5$ Gm, the transfer frequency is $f_*=19.09$ mHz.

The GW response function for relative frequency fluctuations in equal arm Michelson interferometers, which have the data combinations $\eta_{13} + D_{13}\eta_{31} - (\eta_{12} + D_{12}\eta_{21}) $ plus cyclic permutations, is~\cite{Lu:2019log,Schmitz:2020rag}
\begin{equation}
\begin{split}
 \mathcal{R}(u,\gamma) &= s_{2u}\left[s_{\gamma/2}^2\left(\frac{1}{u}+\frac{2}{u^3}\right)+c_{\gamma/2}^2\left(2\,\mathrm{Si}(2u)-\mathrm{Si}(2u_+)-\mathrm{Si}(2u_-)\right)\right]\\
                            &+ c_{2u}\left[s_{\gamma/2}^2\left(\frac{1}{6}- \frac{2}{u^2}\right)+c_{\gamma/2}^2\left(2\,\mathrm{Ci}(2u)-\mathrm{Ci}(2u_+)-\mathrm{Ci}(2u_-)+\ln c_{\gamma/2}^2 \right)\right]\\
                           &- \frac{s_{u_+-u_-}}{32u\;s_{\gamma/2}^3}\left(21-28c_{\gamma}+7c_{2\gamma}+\frac{3-c_{\gamma}}{u^2}\right)+\frac{c_{u_+-u_-}}{8u^2s_{\gamma/2}^2}\left(1+s_{\gamma/2}^2\right)\\
  &-2s_{\gamma/2}^2\left(\mathrm{Ci}(2u)-\mathrm{Ci}(u_+-u_-)-\ln s_{\gamma/2}\right)+\frac{3-c_{\gamma}}{12}-\frac{1-c_{\gamma}}{u^2}, \label{Xresponse2}
\end{split}
\end{equation}
where $u=f/f_*$, $s_x=\sin(x)$, $c_x=\cos(x)$, $u_{\pm}=u \pm  u\sin(\gamma/2)$,  $\gamma=\pi/3$ is the opening angle of the laser arms, and Si and Ci are the sine and cosine integrals.
The Michelson $X$ TDI combination introduces a factor $16 \sin^2(f/f_*)$ from the extra time delays. The final power spectral density of $X_2$ for the stochastic gravitational wave signals from galactic binaries and a phase transition is then 
\begin{equation}
S^\text{signal}_{X_2}(f) = 64 \sin^2(2f/f_*)\sin^2(f/f_*)
\mathcal{R}(f/f_*)
\left[S_{\mathrm{gb}}(f) + S_{\mathrm{pt}}(f)\right].
\label{Xsignal}
\end{equation}
At low frequencies, the $X_2$ transfer function for gravitational waves behaves as $(f/f_*)^6$. As for the instrument noise, again assuming constant and equal arm length, our two noise components from Eqs.~\ref{Sacc} and \ref{OMS} enter the full noise power spectral density for TDI $X_2$ as
\begin{equation}
S^\text{noise}_{X_2}(f) = 64 \sin^2(2f/f_*)\sin^2(f/f_*)\left[S_{\mathrm{OMS}}(f) + \left(3+\cos\left(2\omega L\right)\right)S_{\mathrm{Acc}}(f)\right].\label{Xnoise}
\end{equation}
The above is derived from Eq.~\ref{eq:tdi2approx} by inserting a noise-only data stream containing the two dominant noise components~\cite{Babak:2021mhe,Nam:2022rqg}.

In Fig.~\ref{fig:data} we show the contributions to the power spectral density $S_{X_2}(f)$ (top panel) of the LISA instrument noise (blue), the GB confusion noise (green), and an example injected PT model with $\Omega_{\mathrm{pt}}=10^{-11}, \ f_{\mathrm{pt}}=6 \times 10^{-3}$ Hz (yellow). 
The bottom panel of Fig.~\ref{fig:data} shows the time series of TDI $X_2$ for the same data. We can see in the figure that when considering the full frequency range between $10^{-5}$ Hz and $10^{-1}$ Hz, the instrument noise is significantly larger than the other signals. We also show a zoomed version of the time series in Fig.~\ref{fig:data}, where the annual modulation of the GB confusion noise is more apparent.

\section{Data processing}
\label{sec:proc}
Our \texttt{PyTDI} output is a time series of the TDI $X_2$ variable, corresponding to one year of data with a sampling rate $T_\mathrm{samp} = 5\,$s (see Appendix~\ref{sec:pipeline} for more details). We perform two different analyses with this data: one where we consider the full time series as one dataset, and one in which we divide our time domain data into $N_\mathrm{ch}$ chunks in order to recover the modulated foreground. In both cases, we transform the time series into frequency domain for analysis purposes using the weighted overlapped segment averaging (WOSA) method~\cite{1967ITAE...15...70W,Solomon:1991,Percival_Walden_1993}. 
To do this, the time-domain data is divided into $N_\mathrm{w}$ segments to be windowed before the transform. This means multiplying the data $x(t)$ in each segment by a window function $w(t)$,
\begin{eqnarray}
  x_w(t)&\equiv& w(t)x(t)\label{eq:winx}
\end{eqnarray}
where $w(t)$ has a value of 1 at the center of the interval and tapers off at the start and the end, effectively erasing discontinuities in periodic signals. Next, the data is Fourier-transformed in each windowed segment, and finally averaged over the resulting $N_\mathrm{w}$ periodograms. Use of this method reduces spectral leakage to neighbouring frequency bins, which may become a problem when using a plain discrete Fourier transform~\cite{Heinzel:2002}.  

We choose the popular combination of a Hann\footnote{Often referred to as Hanning.} window
\begin{eqnarray}
  w(t) \equiv \frac{1}{2}\left(1-\cos(2\pi t/T)\right),\label{eq:hann}
\end{eqnarray}
with a 50\% overlap~\cite{Romano:2016dpx, Lazzarini:2004}. This compensates for the loss of time-domain data resulting from the attenuation of the signal at the start and end of each segment, as every data point is effectively counted twice. The Welch estimate of the power spectral densities is carried out using the \texttt{welch} function in the SciPy signal processing toolbox~\cite{2020SciPy-NMeth}.

\begin{figure}[t]
  \centering
  \includegraphics[width=0.95\textwidth]{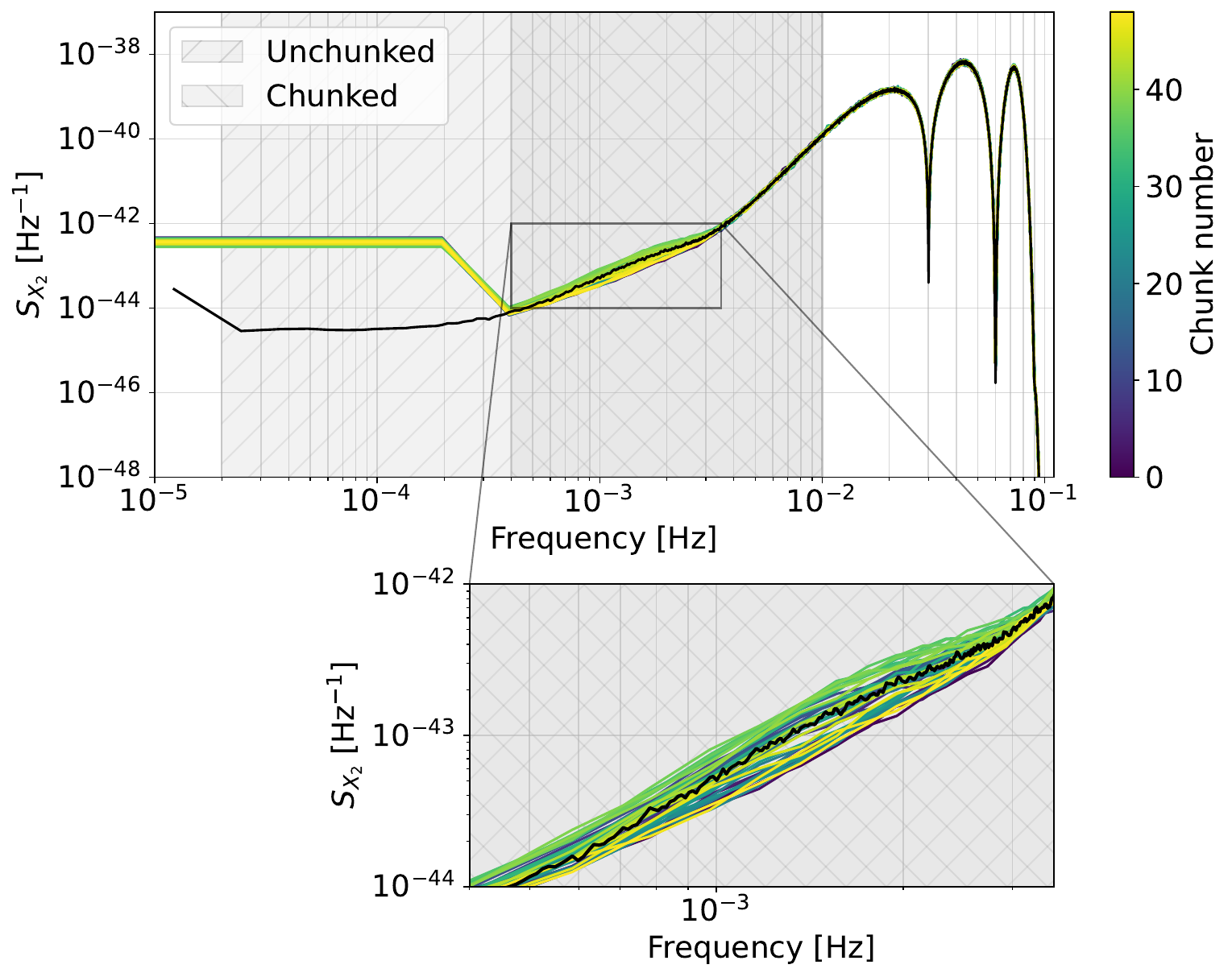}
  \caption{Total power spectral density of the full dataset (black) and of each chunk (colour gradient). The grey hatched regions show the frequency limits for the MCMCs on both the chunked and unchunked data. The bottom panel shows a zoomed in version of the region where the annual modulation has the biggest impact.}
  \label{fig:chvsunch}
\end{figure}

As mentioned above, we process our TDI data in two different ways to carry out two different analyses. First, we transform the time-domain data in one go, setting the segment length to $N_\mathrm{s} = 16 \times 1024$. This will yield the average power spectral density, and we will end up losing the modulation information of the GB foreground. Second, for the purposes of recovering the modulated foreground, we start by dividing our time domain data into $N_\mathrm{ch}$ chunks, where we take $N_\mathrm{ch} = 48$, which is chosen to be close to the value used in Ref.~\cite{Adams:2013qma}. In each chunk, the amplitude may be approximated as constant. We then apply Welch's method with a segment length of $N_\mathrm{s} = 1024$ to the data in each chunk. In Fig.~\ref{fig:chvsunch} we show the power spectral density of the whole data without chunking (black) and the power spectral densities in each of the 48 chunks (colour gradient).

For the data analysis, we will not take the full frequency interval given by the Welch transform, but rather focus on the range our injected models lie in. At frequencies above $f=10^{-2}$ Hz, we mostly have noise, so we set that as our upper limit. The number of data points for the transform defines our lower frequency cutoff. With $N_\mathrm{s} = 16 \times 1024$, we take a lower cutoff of $f=2\times 10^{-5}$ Hz. With $N_\mathrm{s} = 1024$, which we choose for the chunked datasets, the loss of information begins at higher frequencies, as can be seen in Fig.~\ref{fig:chvsunch}, and so we take a lower cutoff of $f=4\times 10^{-4}$ Hz. Thus in the chunked cases, our available frequency interval for the data analysis is narrower.

Note that with the window sizes mentioned above, the number of windows in each of the 48 time chunks is larger than the number of windows for the full time series. Namely, when dividing the data into 48 chunks, we have $N_\text{w}=255$ overlapping segments in each, while in the full time series transform, we end up with $N_\text{w}=769$ segments. In general, increasing $N_\text{w}$ (or decreasing $N_\text{s}$) leads to a lower variance, but also results in more of the aforementioned loss of low-frequency data. The latter effect is especially apparent when we have a low number of data points in each of the 48 chunks. Our current choice of $N_\text{s}=1024$ for $N_\text{ch}=48$ results in a compromise between these two effects.

\section{Likelihood and model comparison}
\label{sec:stats}
\subsection{Likelihood for averaged spectra}
\paragraph{General case:} Consider a time series of duration $T_\text{obs}$ with a sample interval $T_\text{samp}$, yielding a time series of $L = T_\text{obs}/T_\text{samp}$ data points $X_2(t)$. 
For now, we consider the simplest case, meaning data without chunking, without overlaps in the segments, and without applying a window function to the segments. This corresponds to the method of averaged periodograms, also known as Bartlett's method.
We therefore start with $N_\mathrm{w}$ non-overlapping segments containing 
$L_\text{w} = L/N_\mathrm{w}$ data points, and assume that $L_\text{w}$ is an integer. 
The Fourier transform of the $i$th segment $\tilde{X}^i_2(f_n)$ contains 
frequencies $f_n = n/L_\text{w}T_\text{samp} $ with $ 0 \le n <  L_\text{w}$. 
Frequencies with $ L_\text{w}/2 < n$ are complex conjugates of those with $ L_\text{w} - n $, 
hence there are $N_\mathrm{f} = L_\text{w}/2$ independent frequencies. 
The one-side power spectral density $P^i(f_n)$ of the $i$th segment is 
\begin{equation}
  P^i(f_n) = 2|\tilde{X}^i_2(f_n)|^2,
  \label{eq:psdcomps}
\end{equation}
with $0 \le n < L_\text{w}/2$.

We define $\overline{P}(f_n)$ as the average over the segments of the power spectral densities,
\begin{equation}
  \overline{P}(f_n) = \frac{1}{N_\mathrm{w}} \sum_{i=1}^{N_\mathrm{w}} P^i(f_n) \, .
  \label{eq:avg_psd}
\end{equation}

As this is the average of the squares of $2N_\mathrm{w}$ independent standard normal variables, $\overline{P}(f_n)$ follows a chi-squared distribution. Therefore, the 
likelihood function for a model with power spectral density $S(f_n)$ is given by 
\begin{equation}
\mathcal{L} \left(\overline{P} \mid S \right)=\prod_{n=0}^{N_\mathrm{f}-1} \frac{1}{2^{(\nu/2)}\Gamma(\nu/2)}
\left(\frac{\nu}{S(f_n)}\right)
\left(\nu \frac{\overline{P}(f_n)}{S(f_n)}\right)^{(\nu/2)-1} \exp \left(-\frac{\nu}{2} \frac{\overline{P}(f_n)}{S(f_n)}\right) \, ,
  \label{eq:like}
\end{equation}
where $\nu$ is the number of degrees of freedom of the chi-squared distribution, equal to $2N_\mathrm{w}$. 

Similar expressions for the likelihood of averaged spectra appear in Refs.~\cite{Pieroni:2020rob, Gowling:2021gcy},  where the likelihood is then simplified, using the central limit theorem, to a Gaussian distribution for $\overline{P}(f_n)/S(f_n)$. Here we instead use the full description of the likelihood as given by Eq.~\ref{eq:like}.

\paragraph{Specialising to overlapping windowed data:} 
Welch's method improves on the method of averaged periodograms by introducing a window function to reduce spectral leakage, and overlaps to compensate for the loss of information. 
When using Welch's method and a Hann window with a $50\%$ overlap and with windows of length $L_\text{w}$, the number of windows is $N_\text{w} = 2L/L_\text{w} - 1$.
The overlaps introduce correlations between $P^i(f_n)$ from different segments; nevertheless the probability distribution for the mean $\overline{P}(f_n)$ is well approximated by a chi-squared distribution, with an adjusted degrees of freedom parameter.  
To obtain the degrees of freedom for this case, we follow the calculation done in Refs.~\cite{Percival_Walden_1993, Solomon:1991}\footnote{We note that in the last equation on page 29 of Ref.~\cite{Solomon:1991} there is a missing factor 2 in the denominator, which does appear in the similar equation 292b of Ref.~\cite{Percival_Walden_1993}.}, which gives 
\begin{equation}
{\nu}=(36/19)\frac{{N_\mathrm{w}}^2}{{N}_\mathrm{w}-1}, \quad (\text{50\% overlap, Hann window}). 
\label{eq:nuHann}
\end{equation}

\subsection{MCMC methodology for modulated LISA data}
We now apply the above general likelihood to the case of our simulated and processed LISA data.
As discussed in Section~\ref{sec:proc}, we consider a subset $F$ of the full set of frequencies, so that the total number of frequencies is less than $L_\text{w}$. 
Therefore, our final likelihood function will be given by Eq.~\eqref{eq:like} with 
the product taken over a subset of frequencies $f_n \in F$,  with
${N}_\mathrm{w} = 2L/L_\text{w} - 1$, and $\nu$ as given in Eq.~\eqref{eq:nuHann}.

In order to take into account the annual modulation, we will divide our time-domain data into $N_\text{ch} = 48$ non-overlapping chunks of the same length, 
$L_\text{ch} = L/N_\text{ch}$,  
which are further windowed into segments of length $L_\text{w}$ with a 50\% overlapping Hann window, as described previously. 
In the $k$th chunk, our likelihood $\mathcal{L} \left(\overline{P}_k \mid S_k \right) $ will be given by Eq.~\eqref{eq:like} with 
${N}_\mathrm{w} = 2L_\text{ch}/L_\text{w} - 1$, resulting in the final likelihood for chunked data 
$\{\overline{P}_k\}$,
\begin{equation}
\mathcal{L}_\text{ch}\left( \{\overline{P}_k\} \mid \{S_k\} \right)= \prod_{k=1}^{N_\text{ch}} \mathcal{L} \left(\overline{P}_k \mid S_k \right) \, .
\label{eq:like_chunked}
\end{equation}
where $\{S_k\}$ is the set of model power spectral densities, which includes annual modulation.

We implement these likelihood functions in the parameter extraction code \texttt{cobaya}~\cite{Torrado:2020dgo}, which we use to recover both the injected noise contributions and the injected PT signal via MCMC analyses. Specifically, we use the \texttt{mcmc} sampler in \texttt{cobaya}, which is based on the Metropolis Hastings algorithm as presented in \cite{Lewis:2013hha}. As we do not encounter any multi-modal distributions in our results, this sampler is sufficient for our analyses.

All of our analyses will feature the following base parameters: 
\begin{itemize}
\item Two parameters for the instrument noise: $S_\mathrm{a} =A^2_\mathrm{Acc}$ and $S_\mathrm{p} =A^2_\mathrm{OMS}$.
\item Three parameters describing the white dwarf confusion noise: $A_{\mathrm{gb}}$, $f_{\mathrm{gb}}$, and $\gamma_{\mathrm{gb}}$. We fix $\alpha_\mathrm{gb} = -7/3$ in all our analyses as per the discussion in Section~\ref{sec:data}.
\end{itemize}
These parameters will be kept constant in all of our simulated datasets.\footnote{While the settings and input parameters for the different noise sources are the same, the generated data will be different in each run, as the noise itself is generated from a random seed; however, these noise contributions will share the same statistical properties.} All of our datasets will also include an injected PT signal, given by a combination of $\Omega_{\mathrm{pt}}$ and $f_{\mathrm{pt}}$.
Specifically, we will consider all combinations of
\begin{align*}
  \Omega_{\mathrm{pt}} & \in \{ 6\times 10^{-12}, \,  7.8\times 10^{-12}, \, 1\times 10^{-11}, \, 1.8\times 10^{-11}, \, 3\times 10^{-11}, \, 4.2\times 10^{-11}, \, \\
  & \qquad  6\times 10^{-11}, \, 1\times 10^{-10}, \, 3\times 10^{-10}\}; \\
  f_{\mathrm{pt}}/\mathrm{Hz} & \in \{ 4.2\times10^{-4}, \,  6\times10^{-4}, \, 7.8\times10^{-4}, \, 1\times10^{-3}, \, 1.8\times10^{-3}, \, 3\times10^{-3}, \, \\
    & \qquad 6\times10^{-3}, \, 1\times10^{-2}  \}. 
\end{align*}
When running our MCMCs, we will always vary the base parameters, and we will have the following parameters that will be either varied in the MCMC analyses or set to $0$ in our different analyses:
\begin{itemize}
\item Two parameters describing the PT signal: $\Omega_{\mathrm{pt}}$ and $f_{\mathrm{pt}}$. 
\item Four parameters describing the annual modulation of the galaxies: two amplitudes $A_1$ and $A_2$, and two phases $\phi_1$ and $\phi_2$, as seen in Eq.~\ref{eq:mod}. See Appendix~\ref{sec:modulation_parameters} for more information about this choice.
\end{itemize}

For all parameters, we choose the starting values for the MCMC analyses to be the values injected in the simulated data, and we chose the prior range so that the distributions are approximately symmetric around the starting values. The exact prior ranges for the parameters are listed in Table~\ref{tab:priors}.

\begin{table}[t]
	\centering
\begin{tabular}{ccccc} \hline
  & \textbf{Prior} & \textbf{Starting value} & \textbf{Minimum} & \textbf{Maximum} \\ \hline
  $\log_{10} S_\mathrm{a}$ & logarithmic & -29.05 & -23.24 & -34.85 \\ 
  $\log_{10} S_\mathrm{p}$ & logarithmic & -21.65 & -17.32 & -25.98 \\ 
  $\log_{10}A_{\mathrm{gb}}$ & logarithmic & -44.85 & -48 & -42 \\ 
  $\log_{10}f_{\mathrm{gb}}$ & logarithmic & -2.89 & -3.30 & -2.52\\    
  $\gamma_{\mathrm{gb}}$ & linear & 900 & 500 & 1200 \\
  $\log_{10}\Omega_{\mathrm{pt}}$ & logarithmic & model specific & starting value -1 & starting value +1 \\
  $\log_{10}f_{\mathrm{pt}}$ & logarithmic & model specific & starting value -1 & starting value +1 \\
  $A_1$ & linear & 0.1 & 0 & 1 \\
  $A_2$ & linear & 0.5 & 0 & 1 \\
  $\phi_1$, $\phi_2$ & linear & $\pi$ & 0 & 2$\pi$\\ \hline
	\end{tabular}
	\caption{Prior type, starting value, and prior range for all of the parameters that can be varied in our different MCMC analyses.}
	\label{tab:priors}
\end{table}

\begin{table}[t]
	\centering
	\begin{tabular}{cccc} \hline
	\multirow{2}{*}{\textbf{Label}} & \multicolumn{2}{c}{MCMC parameters} & \multirow{2}{*}{\textbf{Chunking}} \\
  & \textbf{Modulation} & \textbf{Phase transition} & \\ \hline
	0 & No & No & No \\ 
	P & No & Yes & No \\ 
	Pc & No & Yes & Yes \\ 
	Mc & Yes & No & Yes \\ 
	MPc & Yes & Yes & Yes \\ \hline
	\end{tabular}
	\caption{Summary of the different MCMC parameters based on whether we are including modulation (M) and PT (P) parameters in our analyses, and whether we are chunking (c) the data.}
	\label{tab:mcmcs}
\end{table}

Finally, when taking into account the annual modulation, we will divide our time series data into $N_\mathrm{ch}$ chunks (before applying Welch's method). Hence, we will have several different models we fit to the data (summarised in Table~\ref{tab:mcmcs}), depending on whether we are including modulation (M) and PT (P) parameters in our analyses, and whether we are chunking (c) the data. In each of these set-ups, we run a grid of MCMCs over datasets featuring different combinations of injected PT signals $\Omega_{\mathrm{pt}}$. 

In order to assess the detectability of our model, we use the Deviance Information Criteria (DIC)~\cite{Spiegelhalter:2002yvw,Spiegelhalter:2014}, 
which can be easily calculated from the posterior distributions obtained from MCMC analyses.
The DIC includes a penalisation term, which penalises over-fitting the model.
Given some data $y$, a model with parameters $\theta$, a posterior mean of $\overline{\theta}$, and deviance $D(\theta) = -2\log\mathcal{L}(y|\theta)$ will have a DIC of
\begin{equation}
\text{DIC} = D(\overline{\theta}) + 2p_D \,.
\end{equation}

Here $p_D$ is the penalisation term, which we take to be $p_D = \overline{D(\theta)}-D(\overline{\theta})$, as done in Ref.~\cite{Gowling:2021gcy}.
The larger the number of parameters, the easier it is for the model to fit the data, so the deviance is penalised. All of these quantities can be calculated from an MCMC sample of the posterior distribution. 

We can use the difference in DIC between two models with a different number of fit parameters to assess which model provides a better fit to the data. In practice, this means running two MCMCs on each dataset with a specific injected fiducial model, with and without the parameters in question, thus evaluating how much the inclusion of the chosen parameters in the MCMC affects the overall DIC.
A higher $\Delta$DIC indicates that the inclusion of the parameters leads to an overall better fit than the null hypothesis.
We note, however, that a DIC comparison is strictly applicable only in the case where the posteriors follow Gaussian distributions, and in cases where the two compared scenarios lead to similar outcomes.

\section{Results}
\label{sec:results}

\paragraph{PT recovery without annual modulation: }
The first question we want to address concerns the detectability of different PT signals when we do not take into account the annual modulation. To address this, we run 72 different MCMCs on data in which we have injected a PT signal given by a combination of $\left\lbrace \Omega_\mathrm{pt} , f_\mathrm{pt} \right\rbrace$, for scenarios 0 and P. We then compare the DICs between the two scenarios with the same $\left\lbrace \Omega_{\mathrm{pt}} , f_{\mathrm{pt}} \right\rbrace$. The results of this analysis are shown in Fig.~\ref{fig:scatter_0vsP}. For numerical values corresponding to the grid points, see Appendix~\ref{sec:dDICvalues}.

\begin{figure}[t]
  \centering
  \includegraphics[width=0.8\textwidth]{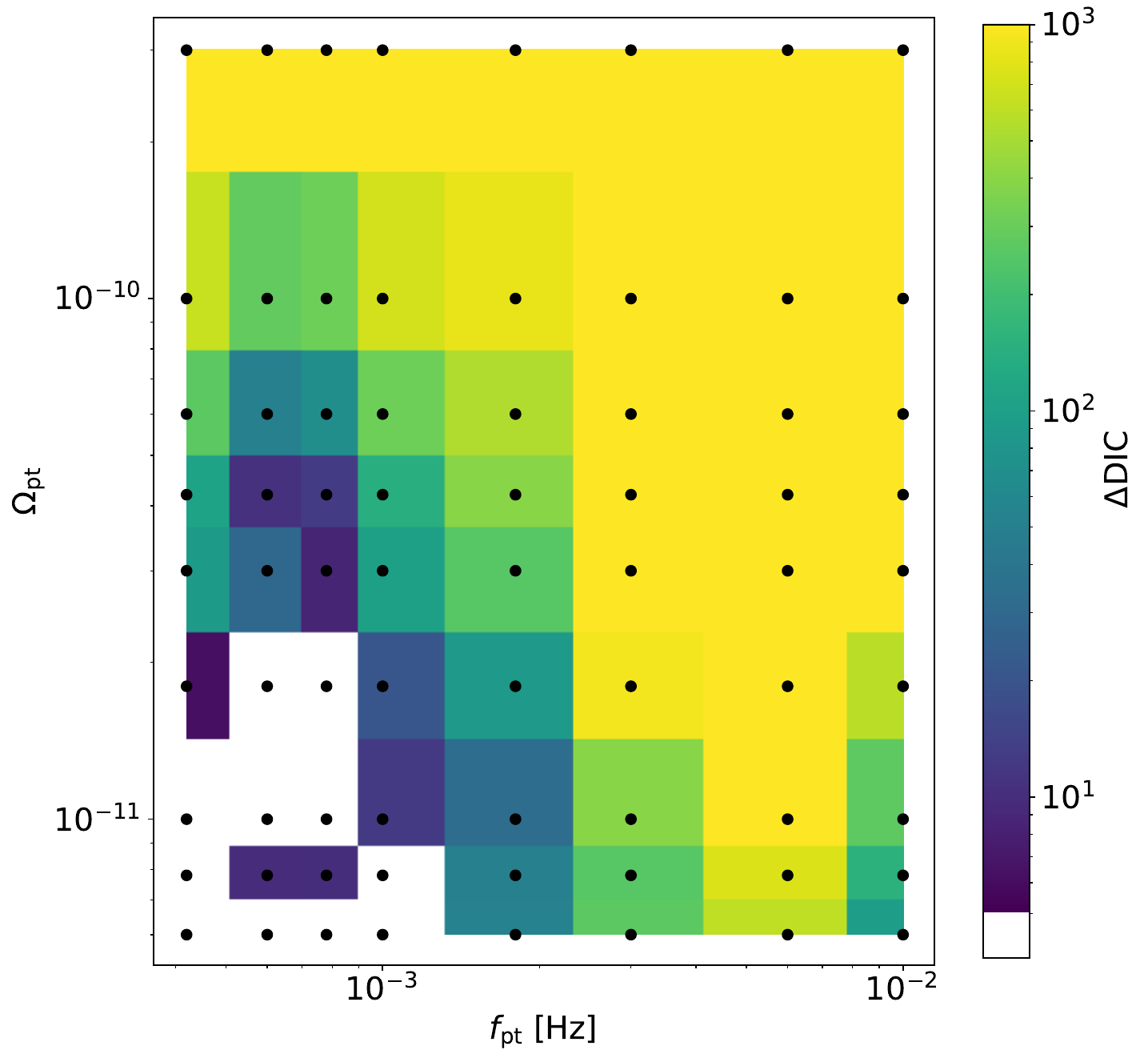}
  \caption{$\Delta$DIC as a function of an injected amplitude $\Omega_\mathrm{pt}$ and peak frequency $f_\mathrm{pt}$, for models with a PT signal (P) compared to the case with no injected PT signal (0), without chunking or considering annual modulation. The colours show the $\Delta$DIC values (filled with a nearest-neighbours interpolation), while the dots indicate the values $\{\Omega_\mathrm{pt},f_\mathrm{pt}\}$ injected in the data. Yellow indicates $\Delta\mathrm{DIC}>1000$, and white indicates $\Delta\mathrm{DIC}<5$. 
  }
  \label{fig:scatter_0vsP}
\end{figure}

We divide the results into three categories. 
In the first category, the model without the PT parameters in the analysis has a DIC very close ($\Delta \mathrm{DIC} < 5$) to that of the model with the PT parameters (white fill colour in Fig.~\ref{fig:scatter_0vsP}), which indicates that the inclusion of the PT parameters does not improve the goodness-of-fit. In the second category, corresponding to models with higher values of $\left\lbrace \Omega_\mathrm{pt} , f_\mathrm{pt} \right\rbrace$, the inclusion of the PT parameters in the analysis significantly improves the goodness-of-fit, leading to $\Delta \mathrm{DIC} > 1000$ (yellow fill colour in Fig.~\ref{fig:scatter_0vsP}). Finally, the category in between these two extremes (marked with colours between yellow and dark blue in Fig.~\ref{fig:scatter_0vsP}) is the one where the DIC becomes most relevant, with the $\Delta$DIC gradually increasing for increasing values of $\Omega_\mathrm{pt}$ and as $f_\mathrm{pt}$ approaches the peak sensitivity frequency around $6$ mHz.

When not taking into account the annual modulation of the galactic binaries, we find that we can accurately recover the injected PT signal for most models with either $\Omega_\mathrm{pt} \geq 10^{-10}$ or $f_\mathrm{pt} \geq 3\times 10^{-3} \,\mathrm{Hz}$, although models below $\Omega_\mathrm{pt} = 2\times 10^{-11}$ become harder to recover regardless of the frequency. 
These results are consistent with Ref.~\cite{Boileau:2022ter}, despite the different model of the galactic binary foreground.

\paragraph{PT recovery with annual modulation: }
Next we want to assess the detectability of different PT signals when we take into account the annual modulation of the galactic binaries. As before, we run 72 different MCMCs on data in which we have injected a PT signal given by a combination of $\left\lbrace \Omega_\mathrm{pt} , f_\mathrm{pt} \right\rbrace$, for scenarios Mc and MPc. We once again compare the DICs between the two scenarios with the same $\left\lbrace \Omega_\mathrm{pt} , f_\mathrm{pt} \right\rbrace$. The results of this analysis are shown in Fig.~\ref{fig:scatter_McvsMPc}, and the corresponding numerical values can be found in Appendix~\ref{sec:dDICvalues}.

\begin{figure}[t]
  \centering
  \includegraphics[width=0.8\textwidth]{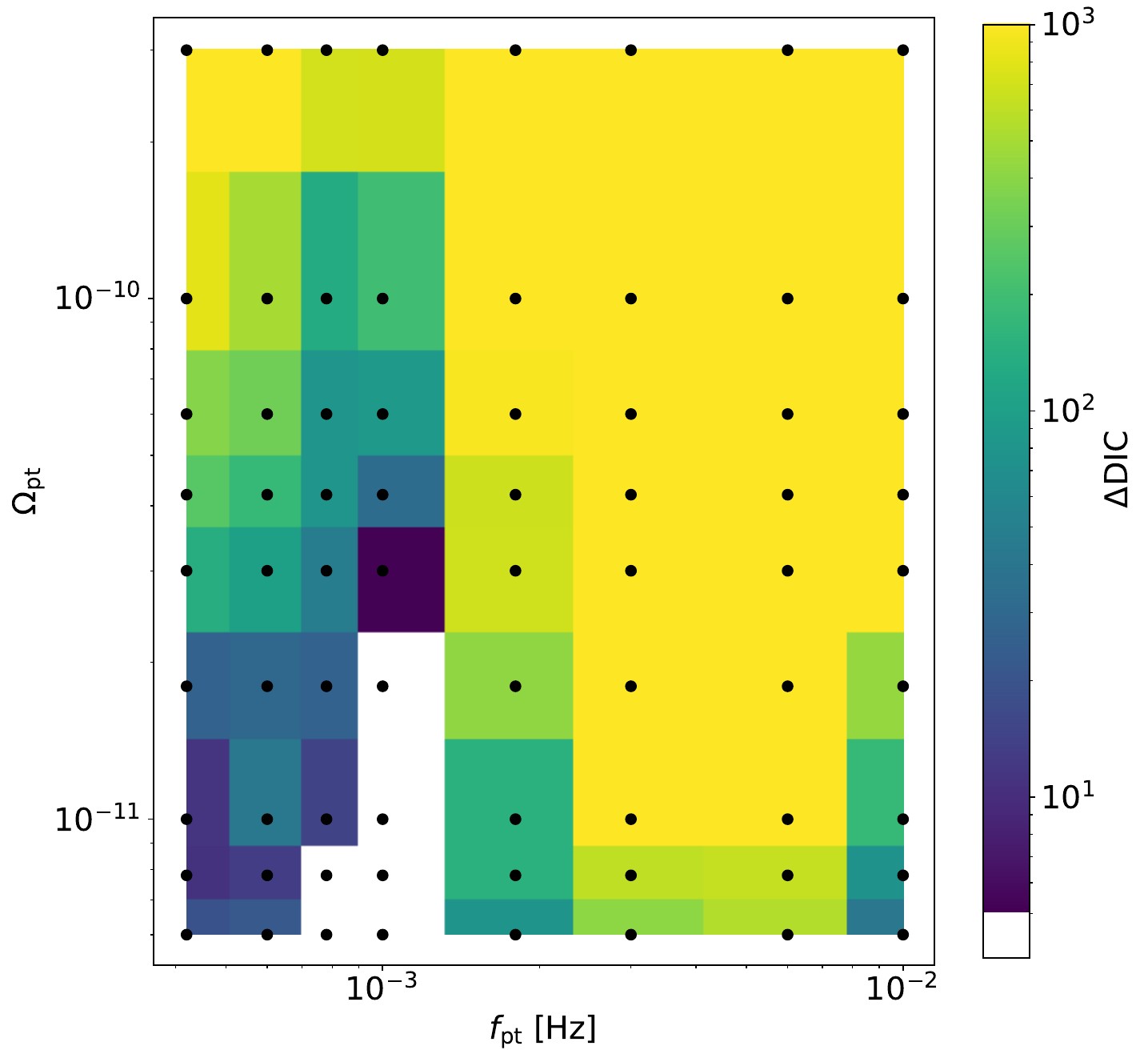}
  \caption{$\Delta$DIC as a function of an injected amplitude $\Omega_\mathrm{pt}$ and peak frequency $f_\mathrm{pt}$, for models with a PT signal (MPc) compared to the case with no injected PT signal (Mc), when chunking and considering annual modulation. The colour gradient shows the $\Delta$DIC values (filled with a nearest-neighbours interpolation), while the dots indicate the values $\{\Omega_\mathrm{pt},f_\mathrm{pt}\}$ injected in the data. Yellow indicates $\Delta\mathrm{DIC}>1000$, and white indicates $\Delta\mathrm{DIC}<5$.}
  \label{fig:scatter_McvsMPc}
\end{figure}

We can once again divide the results into three different categories, as we did in Fig.~\ref{fig:scatter_0vsP}: cases where the inclusion of the PT parameters does not improve the goodness-of-fit (white in Fig.~\ref{fig:scatter_McvsMPc}); cases where the inclusion of the PT parameters in the analysis significantly improves the goodness-of-fit, leading to $\Delta \mathrm{DIC} > 1000$ (yellow in Fig.~\ref{fig:scatter_McvsMPc}); and scenarios in between these two extremes (colours between yellow and dark blue in Fig.~\ref{fig:scatter_McvsMPc}).

In summary, when exploiting the annual modulation of the galactic binaries, we find that we can accurately recover the injected PT signal for all models with $\Omega_\mathrm{pt} \geq 3\times 10^{-11}$ and $f_\mathrm{pt} \geq 2\times 10^{-3} \,\mathrm{Hz}$, as well as some models with lower amplitudes (down to $\Omega_\mathrm{pt} \sim 10^{-11}$) in the frequency range $2\times 10^{-3} \,\mathrm{Hz} < f_\mathrm{pt} \leq 7\times 10^{-3} \,\mathrm{Hz}$.  For $f_\mathrm{pt} = 6\times 10^{-3} \,\mathrm{Hz}$ we can recover all injected PT signals, down to our lowest injected amplitude of  $\Omega_\mathrm{pt} = 6 \times 10^{-12}$.
We note that, comparing to the flat spectrum recovered in Ref.~\cite{Adams:2013qma} in the presence of anisotropic GB confusion noise, our weakest recovered signal here is more than an order of magnitude smaller than what was recovered there. However, in the most sensitive frequency range we have $\Delta\mathrm{DIC}\sim 500$, so we can anticipate that the PT signal will be observable at lower $\Omega_\mathrm{pt}$. It is also to be noted that their instrument noise power was an order of magnitude smaller than in our noise model, as can be seen by comparing their Fig.~1 with our Fig.~\ref{fig:data}.
In general, by comparing Figs.~\ref{fig:scatter_0vsP} and~\ref{fig:scatter_McvsMPc}, we can see that the inclusion of the annual modulation leads to overall higher $\Delta$DIC, and therefore increases the detectability of the PT signal. A notable exception to this is for models with $f_\mathrm{pt} = 10^{-3}\,\mathrm{Hz}$, where the goodness-of-fit decreases when taking into account the annual modulation. This case is discussed in further detail below.

\begin{figure}[t]
  \centering
  \includegraphics[width=1.0\textwidth]{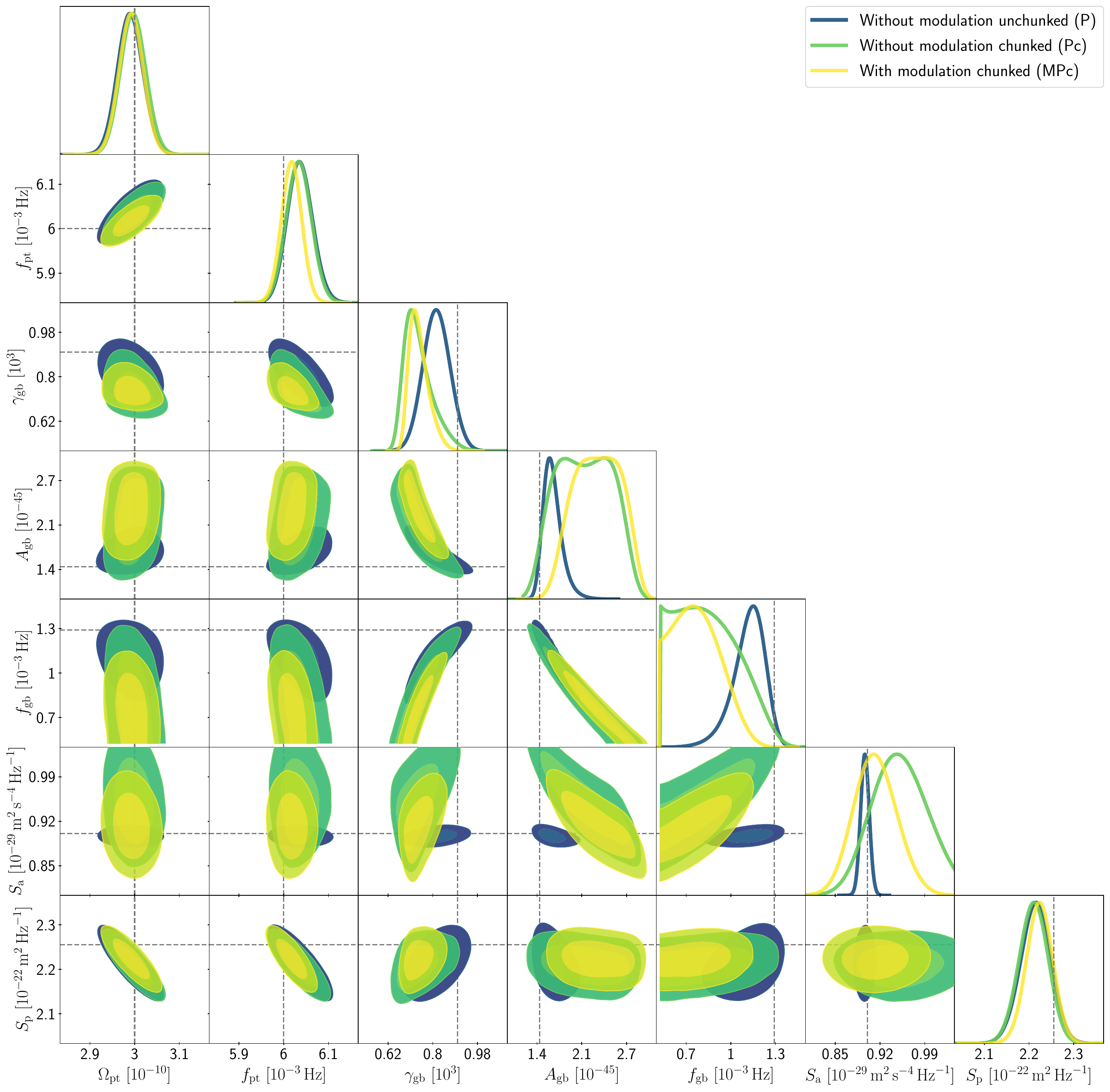}
  \caption{2D posterior distribution showing the $1\sigma$ and $2\sigma$ contours, for an injected signal of $\Omega_{\mathrm{pt}}=3\times10^{-10}, \ f_{\mathrm{pt}}=6\times 10^{-3}$ Hz. The injected values are marked with dashed lines. The blue contours correspond to the case with no chunking or modulation, the green contours indicate chunked data without annual modulations, and the yellow contours show the case where we have chunked the data and included the annual modulation.}
  \label{fig:MCMCbest}
\end{figure}

\paragraph{Impact of chunking and annual modulation on parameter recovery: }
In order to account for the annual modulation, we first divide the data into chunks before applying Welch's method, as described in Section~\ref{sec:proc}. To do this, we need to shorten the segment length for windowing, and this leads to some information loss on low frequencies. Here we aim to assess the impact of this low-frequency loss, and the subsequent inclusion of the annual modulation, on the recovery of the injected signals. 

\begin{figure}[t]
  \centering
  \includegraphics[width=1.0\textwidth]{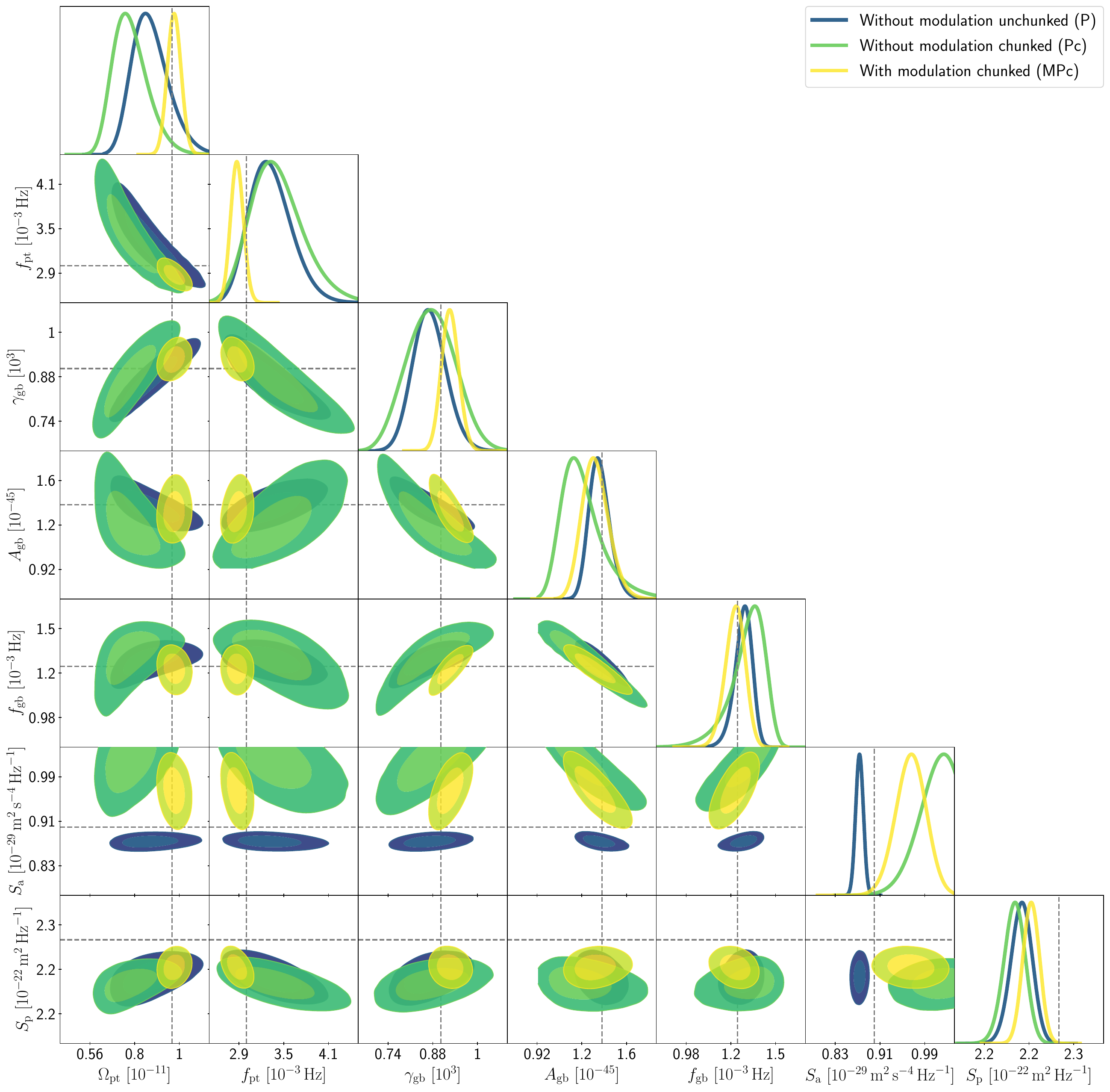}
  \caption{2D posterior distribution showing the $1\sigma$ and $2\sigma$ contours, for an injected signal of $\Omega_{\mathrm{pt}}=10^{-11}, \ f_{\mathrm{pt}}=3\times 10^{-3}$ Hz. The injected values are marked with dashed lines. The blue contours correspond to the case with no chunking or modulation, the green contours indicate chunked data without annual modulations, and the yellow contours show the case where we have chunked the data and included the annual modulation.}
  \label{fig:MCMCinb}
\end{figure}

To see how the chunking and inclusion of the annual modulation affects the parameter recovery, in Fig.~\ref{fig:MCMCbest} we show the posterior distributions obtained from the MCMCs for an injected PT signal with $\Omega_{\mathrm{pt}}=3\times10^{-10}, \ f_{\mathrm{pt}}=6\times 10^{-3}$, while in Fig.~\ref{fig:MCMCinb} we show the same for the case with $\Omega_{\mathrm{pt}}=10^{-11}, \ f_{\mathrm{pt}}=3\times 10^{-3}$. The former corresponds to our strongest PT signal, while the latter corresponds to one of the in-between, weaker, scenarios. In both of these figures, the blue contours correspond to the case with no chunking or modulation (P), the green contours indicate chunked data without annual modulation (Pc), and the yellow contours show the case where we have chunked the data and included the annual modulation parameters in our analyses (MPc).

We also show  the injected values for the parameters with dashed lines in Figs.~\ref{fig:MCMCbest} and~\ref{fig:MCMCinb}. We expect a small shift in the recovered values with respect to the injected ones, due to the small bias that is introduced when using the WOSA method~\cite{SCHOUKENS200627, ANTONI20071723, 2023arXiv231213643A}. Even when factoring this in, we can see that in the unchunked and unmodulated case, we recover the injected values for all the parameters to within $1\sigma$, with the exception of the instrument noise parameters $S_\mathrm{a}$ and $S_\mathrm{a}$, which are slightly shifted when we go to weak injected PT signals, as seen in Fig.~\ref{fig:MCMCinb}. When chunking the data, for some of the GB and the instrument noise parameters we lose accuracy in the signal recovery, leading to broader posteriors, due to the loss of low frequency information. This is especially apparent for the acceleration noise parameter, $S_\mathrm{a}$, which dominates at low frequencies. Otherwise, we can see that the chunking itself does not significantly affect the parameter recovery.

Furthermore, we can see in Fig.~\ref{fig:MCMCbest} that when we have a strong PT signal, the PT signal and the galactic binary confusion noise are already distinct enough that the inclusion of the annual modulation only provides a moderate improvement in the parameter recovery. On the other hand, in Fig.~\ref{fig:MCMCinb}, where we show a much weaker PT signal, we can see that the inclusion of the annual modulation leads to a considerable improvement in the sensitivity to the PT parameters. In this regime, the PT signal and galactic binary confusion noise have similar peak frequencies, so here the inclusion of the annual modulation allows these signals to be more easily separated.

Finally, there is the special case where the PT signal and the galactic binary confusion noise almost the same peak frequencies ($f_\mathrm{pt} \approx f_\mathrm{gb} \approx 10^{-3}\,\mathrm{Hz}$). When this happens, the two signals cannot be disentangled, even when including the annual modulation. In fact, in this scenario, chunking the data removes too much information on low frequencies, which in turn makes it harder to tell the PT signal and galactic binary confusion noise apart. This highlights that while chunking the data to include the annual modulation can help distinguish the PT signal in many cases, it can also lead to too much information loss at low frequencies when the PT signal is very similar to the galactic binary confusion noise. We leave a more detailed analysis on the ideal number of chunks to minimise this information loss for future work.

\section{Conclusions}
\label{sec:conclusions}

Seeing (or confidently constraining) a stochastic background of gravitational waves at future gravitational wave detectors represents a crucial test of new physics beyond the Standard Model. In this paper we have generated data incorporating a broken power law background in the time domain using the LISA Simulation Suite. This allows us to more accurately study how the detector response affects our ability to recover a signal. We have also incorporated a modelled foreground of galactic binaries. Our approach in this paper allows us to generate a signal which exhibits the characteristic annual variation of the galactic binary foreground, and investigate to what extent this affects our ability to recover a phase transition signal.

Given a time series of time delay interferometry data from the LISA Simulation Suite, we then attempt to recover the parameters of both the modelled compact binary foreground and the hypothesised stochastic background coming from a first-order phase transition, and employ the Deviance Information Criterion to determine which model is preferred in each case.

Overall, we have seen that when not considering the annual modulation in our analyses, we can successfully recover PT signals for all models with 
either $\Omega_\mathrm{pt} \geq 10^{-10}$ or $f_\mathrm{pt} \geq 3\times 10^{-3} \,\mathrm{Hz}$, which is compatible with previous results in the literature.
When exploiting the annual modulation of the galactic binaries, we can recover all models with $\Omega_\mathrm{pt} \geq 3\times 10^{-11}$ and $f_\mathrm{pt} \geq 2\times 10^{-3} \,\mathrm{Hz}$, as well as some models with lower amplitudes (down to $\Omega_\mathrm{pt} \sim 10^{-11}$) in the frequency range $2\times 10^{-3} \,\mathrm{Hz} < f_\mathrm{pt} \leq 7\times 10^{-3} \,\mathrm{Hz}$.  For $f_\mathrm{pt} = 6\times 10^{-3} \,\mathrm{Hz}$, we recover all injected models regardless of the amplitude. The inclusion of the annual modulation of the galactic binaries leads to an improvement in most of the $\Delta$DICs. While these results are more conservative than previous estimates in the literature, we note that we are using more up-to-date information about the expected LISA instrument noises than earlier works.

In order to account for the annual modulation, we divide the data into chunks, for which we have to adjust our segment length for the windowing, and this results in a loss of information at low frequencies. 
Despite this, in most cases chunking the data and including the annual modulation leads to an improvement in the goodness-of-fit, and increases our ability to detect a stochastic gravitational wave background. 
However, for models where the frequency of the PT signal is very close to the frequency of the galactic binary confusion noise, the chunking leads to an overall reduction in our ability to distinguish these signals. A more detailed analysis of the impact of the number of chunks and segment length (and therefore the window size) on the parameter recovery is left for future work.

We simulate one year's worth of data in order to be able to fully investigate the effect of the annual variation on LISA. As of adoption, the mission's planned duration is 4 years 
and will be subject to interruptions in the availability of data. In practice, increasing the duration of our simulated data is likely to improve our ability to resolve the PT signals. We have not modelled these effects here. Using the python-based \texttt{LISA Instrument} code was a limiting factor in the duration for which we could generate simulated data, constrained principally by the memory requirement, which increases quickly with the amount of data produced, as described in Ref.~\cite{Bayle:2022okx}.

Although this paper has been concerned with generating and analysing data for the future LISA mission, the problem of model determination for superposed stochastic gravitational wave signals is highly timely: the NANOGrav Pulsar Timing Array collaboration has explored the possibility that new physics may be responsible for discrepancies between their observed signal and the expected background from supermassive black hole binaries~\cite{NANOGrav:2023hvm}. Further work in this area may benefit gravitational wave studies beyond LISA.

\subsection*{Acknowledgements}
We thank Jean-Baptiste Bayle for insightful discussions and feedback.
MH (ORCID ID 0000-0002-9307-437X) was supported by Academy of Finland grant no 333609. 
DCH (ORCID ID 0000-0003-2811-0917) was supported by Academy of Finland grant nos. 328958, 349865 and 353131.
TM (ORCID ID 0009-0005-6855-184X) was supported by Academy of Finland grant no. 349865 and the Magnus Ehrnrooth Foundation.
DJW (ORCID ID 0000-0001-6986-0517) was supported by Academy of Finland grant nos. 324882, 328958, 349865 and 353131.
We acknowledge CSC – IT Center for Science, Finland, for computational resources.

\appendix
\section{Simulating LISA data}
\label{sec:pipeline}
Overview: we use \texttt{LISA Orbits} (v2.3) to generate an orbit file; \texttt{LISA GW Response} (v2.3) to simulate the instrument response to the injected GW signals; \texttt{LISA Instrument} (v1.4) to account for the LISA instrument noise, laser beam propagation, the interferometric measurements, and on-board data processing; and \texttt{pyTDI} (v1.3) to perform the time-delay interferometry calculations. \texttt{LISA Constants} (v1.3.4)~\cite{LISAConstants} provides the necessary physical constants and mission parameters throughout the simulation. At each step in the process, we produce a new data file that is used as input for the next part of the pipeline.

\paragraph{Orbit file:}
In our analysis, we use one full year of data. However, to account for the incomplete orbital information in the first 10 seconds of the orbit file, as well as for the warm-up time of anti-aliasing filters in the instrument simulation, we use \texttt{LISA Orbits} to create orbit files with two years of data. The extra year will also give us leeway if we wish to extend our simulation later. We simulate Keplerian orbits with a time step of 8640 seconds, which is longer than the time step in the rest of the simulation, to prevent the orbit file from being too large. The data will be resampled to a time step of 5 seconds in the next phases of the simulation.

\begin{figure}[t]
  \centering  
  \includegraphics[width=0.45\textwidth]{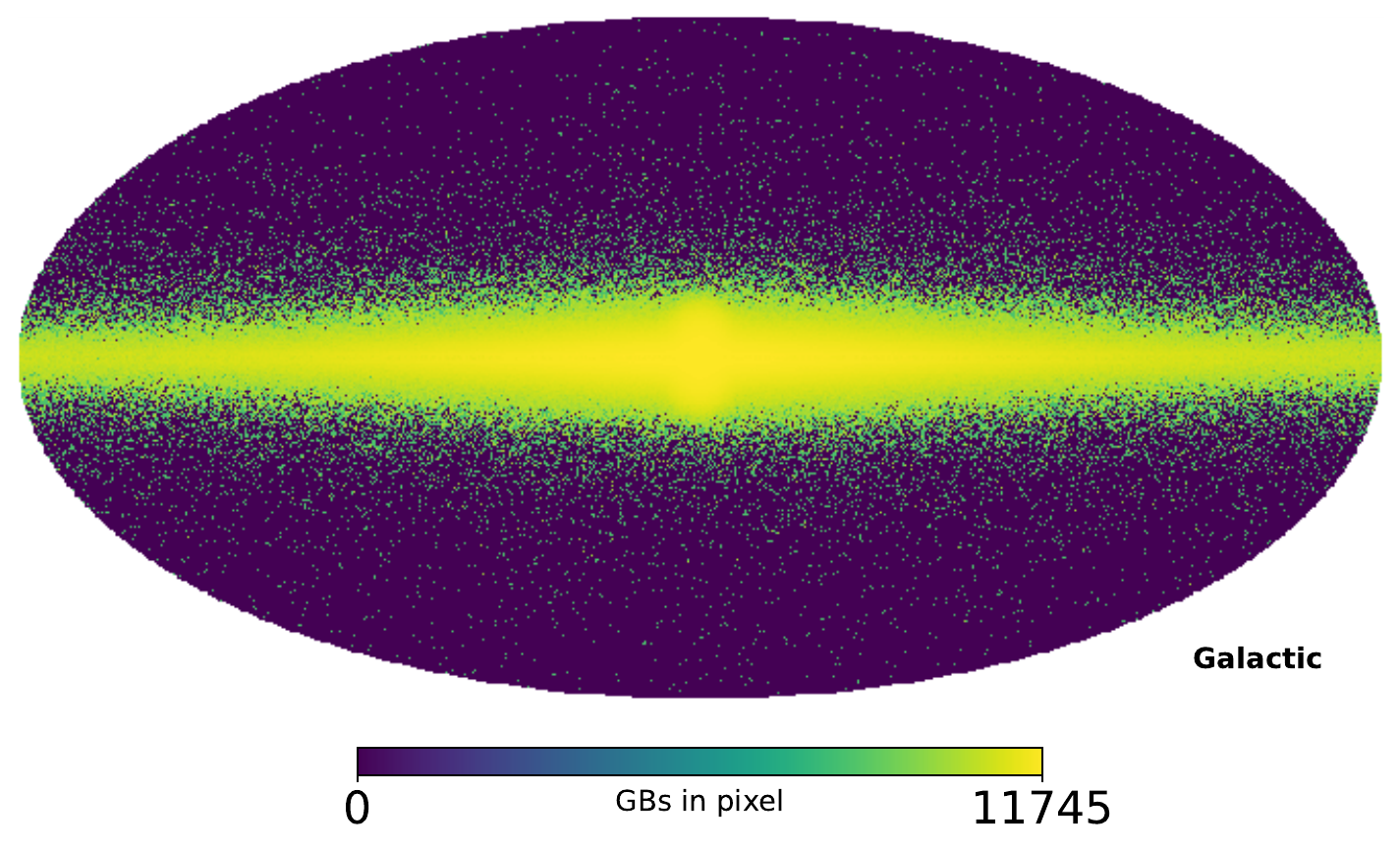}
  \includegraphics[width=0.45\textwidth]{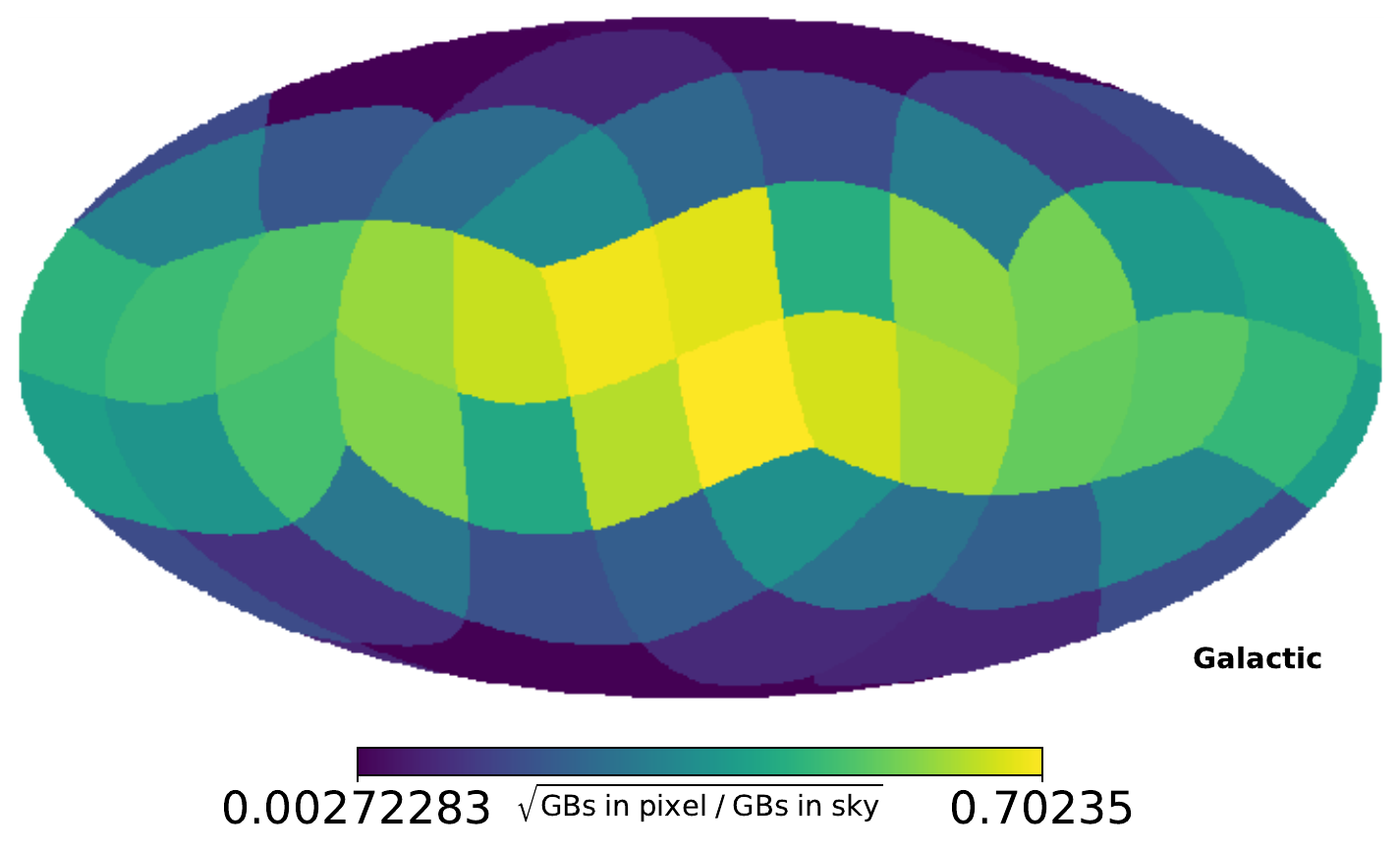}
  \caption{\textbf{Left:} the Sangria binary distribution, with a resolution of 786432 pixels, and the pixel intensity reflecting the total number of binaries in the pixel. \textbf{Right:} \\the skymap we produce with \texttt{HEALPix}, with $N_\mathrm{pix} = 48$.}
  \label{fig:skymap48}
\end{figure}

\paragraph{Galactic binaries:}
In order to simulate the galactic binaries, we start by creating a \texttt{HEALPix} skymap. The angular noise power spectrum for LISA is expected to increase substantially for multipoles $\ell > 6$, as seen in Fig.~13 of Ref.~\cite{LISACosmologyWorkingGroup:2022kbp} for one year of data and Fig.~4 of Ref~\cite{Alonso:2020rar} for four years of LISA data. We can convert this multipole to a corresponding solid angle via $\Omega = 2 (180\deg)^2 / (\pi \ell)$. Following Table 1 of Ref.~\cite{Gorski:2004by}, for the solid angle corresponding to $\ell > 6$ we would need to take a \texttt{HEALPix} skymap with $N_\mathrm{pix} > 12$ pixels. We therefore create a skymap with 48 pixels, into which we insert the coordinates of the GBs contained in the catalogue, such that the skymap pixel intensity is defined by the number of GBs in the pixel normalised by the total number of GBs. Because \texttt{LISA GW Response} requires the square root of the intensity skymap, we further take the square root of our skymap. The resulting skymap is presented in the right panel of Fig.~\ref{fig:skymap48}. A more precise distribution of the binaries entering our skymap is given in the left panel of the same figure.

To inject the GB signal into our simulated data, we call the \texttt{StochasticBackground} class in \texttt{LISA GW Response}, passing it a confusion noise foreground as a function of frequency, along with the anisotropic skymap described above. This effectively inserts the foreground function into each of the skymap pixels, weighting it by the intensity in the pixel. We opt for this approach, instead of using the \texttt{GalacticBinary} class to individually insert each GB, to reduce the computational cost, as we are dealing with millions of binaries.

At this point in our simulation, our time step is 5 seconds. The size of the dataset is 6311900 to allow us to exclude up to 380 points containing possible simulation artifacts. We start the simulation at 10 seconds to ensure we have the full orbital information.

\paragraph{Phase transition:}
To add the PT signal to the data, we once again use the \texttt{Stochastic}\-\texttt{Background} class in \texttt{LISA GW Response}, this time passing an isotropic skymap with pixel intensity set to 1, and a PT signal characterised by a peak amplitude and peak frequency. We take care of the normalization by dividing the peak amplitude by the number of skymap pixels, after which we pass the PT function of frequency to \texttt{StochasticBackground}, along with the simulation parameters listed above.

\paragraph{LISA instrument noise:}
To add the instrument noises to our simulated data, we use \texttt{LISAInstrument}~\cite{Bayle:2022okx}. We pass the GW file from the previous stage to the \texttt{Instrument} class, using the default upsampling with a Kaiser filter for anti-aliasing. We disable laser locking by using the locking configuration \texttt{lock='six'}, so that we get an independent noise for each laser. \texttt{LISA Instrument} provides flexibility to turn different noise contributions on and off. Here we include the dominant noise components, namely the test-mass acceleration noise and the ISI carrier OMS noise, with values set to those defined in the LISA Science Requirements Document~\cite{LISA_SR_doc}. In practice, we set \texttt{testmass\_asds=3e-15, oms\_asds=(15e-12, 0, 0, 0, 0, 0), backlink\_asds=0} and disable all noise categories except for the pathlength noises.

\paragraph{Time-delay interferometry:}
Due to the fluctuating arm lengths in LISA, the signals need to be time-shifted for the laser noise to cancel out as they interfere, with a procedure referred to as time-delay interferometry (TDI). To compute the time delays, we use the publicly available package \texttt{pyTDI}, which provides functions for computing both the first and second generation Michelson combinations $X, Y, Z$, the Sagnac combinations $\alpha, \beta, \gamma$, and the orthogonal combinations $A, E, T$. We use \texttt{pytdi.michelson.X2} to create the 2nd generation Michelson $X$ combination. Since we wish to have the output in fractional frequency deviations, we divide the TDI data by the central frequency of the laser beams $\nu_0=281.6\mathrm{THz}$.

\section{Choosing the number of parameters for the annual modulation}
\label{sec:modulation_parameters}
\begin{figure}[t!]
  \centering
  \includegraphics[width=1\textwidth]{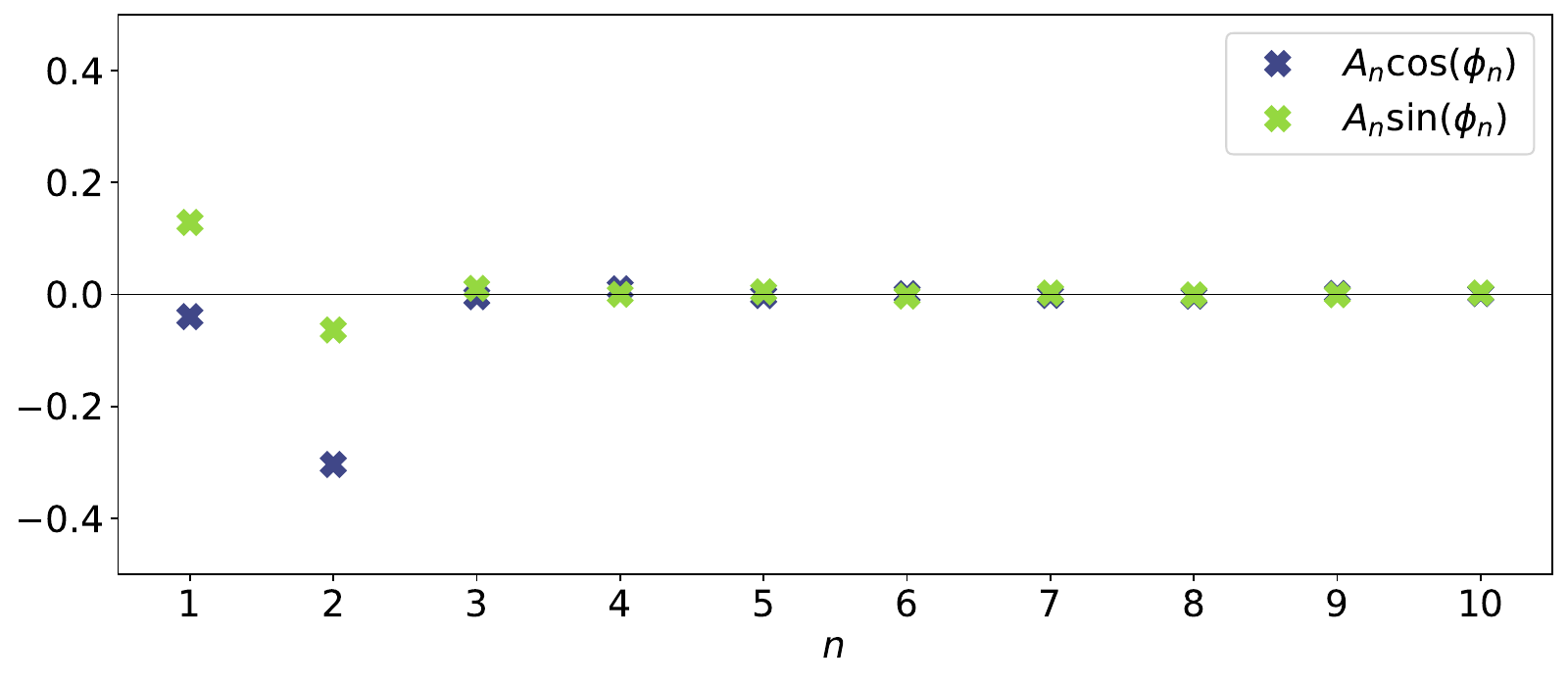}
  \caption{The first 10 real and imaginary Fourier coefficients $A_n\cos(\phi_n)$ and $A_n\sin(\phi_n)$ representing the modulation of the GB foreground, as given in Eq.~\ref{eq:gb_fourier}.}
  \label{fig:fourier}
\end{figure}

First we wish to evaluate how many parameters we need to describe the annual modulation of the GB signal. To do this, we produce a dataset which only contains the GB component, i.e. is free of instrument noise and PT signals, and which can therefore be described by Eq.~\ref{CPSD}. 
As described in Section~\ref{sec:proc}, to account for the annual modulation we divide our data into $N_\mathrm{ch}=48$ time chunks, and we approximate the amplitude of the power spectral density as being constant in each chunk. We then apply the WOSA method to obtain the power spectral density in each chunk, as described in Eq.~\ref{eq:avg_psd}.

Finally, we perform a discrete Fourier transform on the average power spectral density of each chunk (as shown in Eq.~\ref{eq:gb_fourier}) to assess the contribution of each Fourier mode to the total GB signal. We show the amplitude of the real and imaginary Fourier coefficients in Fig.~\ref{fig:fourier}, where we can see that only the first two terms have a noticeable deviation from zero. Similar analyses can be found in~\cite{Ungarelli:2001xu, Adams:2013qma}.

To investigate the statistical difference between including two or three Fourier terms, we perform three MCMCs on the same GB-only data.
In the first MCMC, we do not include any annual modulation parameters in the analysis (model Mc), in the second MCMC we include two terms of the Fourier expansion (M$_2$c), and in the third MCMC we include three terms of the Fourier expansion (M$_3$c). 
By comparing the $\Delta$DICs between these models, we can evaluate if the goodness-of-fit improves when including the various Fourier terms. Between models Mc and M$_2$c, we find $\Delta \mathrm{DIC} \approx 10^{6}$, and between models M$_2$c and M$_3$c we find $\Delta \mathrm{DIC} \approx 10^{2}$. This shows that the inclusion of the first two Fourier terms provides a substantial improvement in the goodness-of-fit, while including the third term only moderately improves the fit further.

\begin{figure}[t]
  \centering  
  \includegraphics[width=1\textwidth]{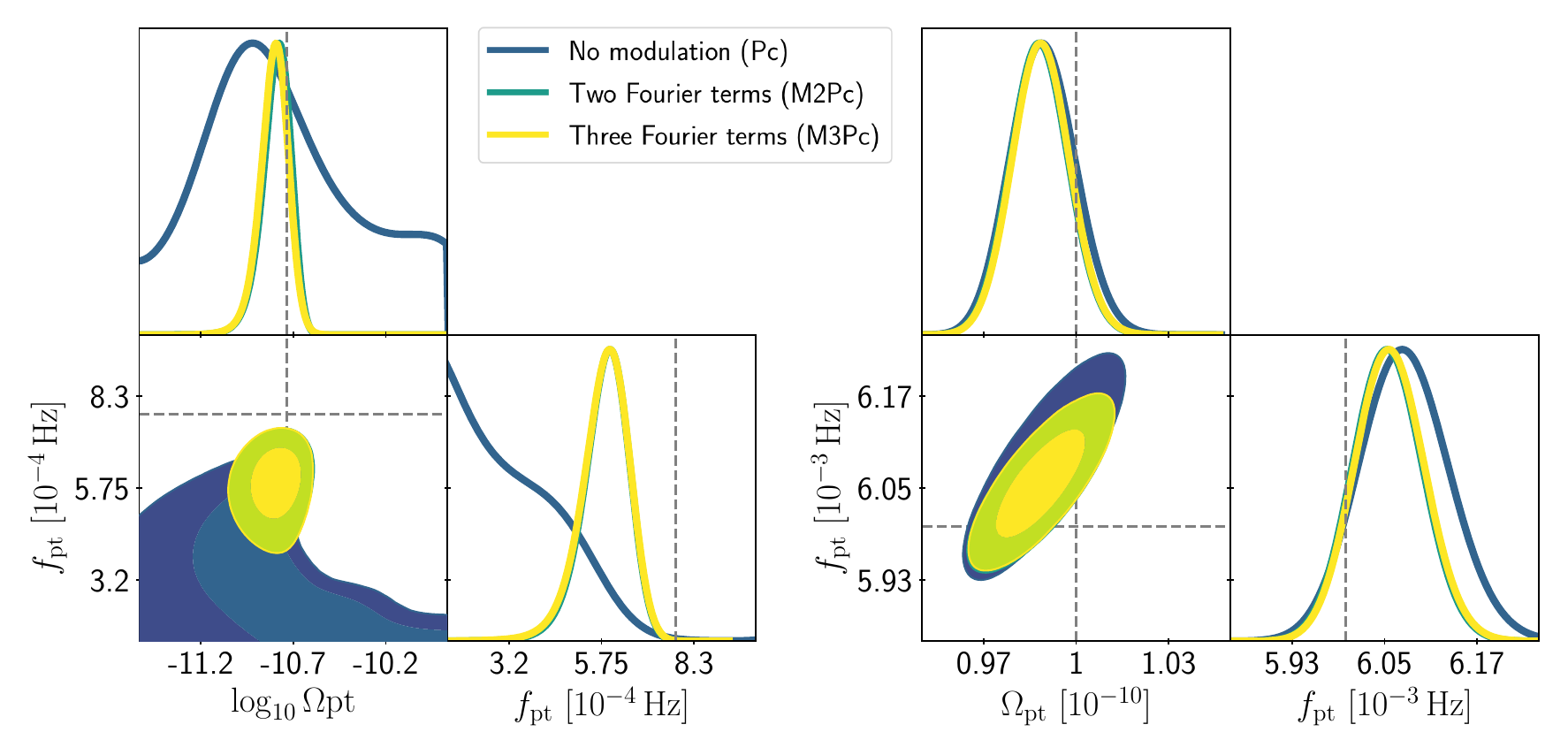}
  \caption{2D posterior distributions showing the $1\sigma$ and $2\sigma$ contours for for $\Omega_{\mathrm{pt}}$ and $f_{\mathrm{pt}}$. The injected values are marked with dashed lines. The blue contours correspond to the case with no modulation (Pc), the green contours indicate the analysis with two Fourier expansion terms (M$_2$Pc), and the yellow contours show the case where we have included three terms in the Fourier expansion (M$_3$Pc). \textbf{Left:} Injected model: $\Omega_{\mathrm{pt}}=1.8\times 10^{-11}, \ f_{\mathrm{pt}}=7.8\times 10^{-4}$ Hz. We note that here we express $\Omega_{\mathrm{pt}}$ in $\log_{10}$ scale, as the contour spans several orders of magnitude. \textbf{Right:} Injected model: $\Omega_{\mathrm{pt}} = 10^{-10}, \ f_{\mathrm{pt}}=6\times 10^{-3}$ Hz. }
  \label{fig:MCMCmod}
\end{figure}

Finally, we investigate the impact of the various modulation terms on the recoverability of the PT parameters. In Fig.~\ref{fig:MCMCmod} we show the results of MCMCs performed on data which includes the GB confusion noise, instrument noise and a PT signal with . In blue we show the case where we do not include any modulation parameters (Pc), in green we show the analysis including two terms of the Fourier expansion (M$_2$Pc), and the yellow contours represent the case where we include three terms of the Fourier expansion (M$_3$Pc). We perform this analysis on data with two different injected PT signals: $\Omega_{\mathrm{pt}}=1.8\times 10^{-11}, \ f_{\mathrm{pt}}=7.8\times 10^{-4}$ Hz (left panel of Fig.~\ref{fig:MCMCmod}) and $\Omega_{\mathrm{pt}} = 10^{-10}, \ f_{\mathrm{pt}}=6\times 10^{-3}$ Hz (right panel of Fig.~\ref{fig:MCMCmod}).
We can see that in both cases, the contours for the M$_2$Pc and M$_3$Pc are almost completely overlapped. In both of these cases, between models Pc and M$_2$Pc, we find $\Delta \mathrm{DIC} \approx 10^{3}$, and between models M$_2$Pc and M$_3$Pc we find $\Delta \mathrm{DIC} \approx 10^{1}$. Overall, we can conclude that including more than two terms in the Fourier expansion does not improve the goodness-of-fit nor the recoverability of the PT parameters.

\section{Tables with $\Delta\mathrm{DIC}$ values}
\label{sec:dDICvalues}
Table~\ref{tab:0vsP} displays $\Delta\mathrm{DIC}$ values corresponding to Fig.~\ref{fig:scatter_0vsP}, where the PT signal (P) is compared to the case with no injected PT signal (0), without chunking or considering annual modulation. Table~\ref{tab:McvsMPc} shows $\Delta\mathrm{DIC}$ values corresponding to Fig.~\ref{fig:scatter_McvsMPc}, where the PT signal (MPc) is compared to the case with no injected PT signal (Mc), with chunking and considering annual modulation.
\begin{table}[t]
  \centering
  \small
  \tabcolsep=0.09cm
  \begin{tabular}{|c| D{.}{.}{4.2} D{.}{.}{4.2} D{.}{.}{4.2} D{.}{.}{4.2} D{.}{.}{4.2} D{.}{.}{6.2} D{.}{.}{6.2} D{.}{.}{6.2} |}
    \hline
    & \multicolumn{8}{c|}{$f_\mathrm{pt}$ [Hz]} \\
    \hline
    $\Omega_\mathrm{pt} $ & \multicolumn{1}{c}{$4.2\times 10^{-4}$} & \multicolumn{1}{c}{$6\times 10^{-4}$} & \multicolumn{1}{c}{$7.8\times 10^{-4}$} & \multicolumn{1}{c}{$1\times 10^{-3}$} & \multicolumn{1}{c}{$1.8\times 10^{-3}$} & \multicolumn{1}{c}{$3\times 10^{-3}$} & \multicolumn{1}{c}{$6\times 10^{-3}$} & \multicolumn{1}{c|}{$1\times 10^{-2}$} \\
    \hline
    $3\times 10^{-10}$ & 3444.18 &  1663.61 &  1741.70 &  2552.27 &  4383.17 &  228815.13 &  690962.06 &  397976.00\\
    $1\times 10^{-10}$ & 637.00 &  279.70 &  307.76 &  700.55 &  842.10 &  49886.02 &  129207.64 &  46621.63\\
    $6\times 10^{-11}$ & 265.05 &  50.49 &  67.88 &  312.75 &  535.98 &  2502.14 &  47732.15 &  9449.26\\
    $4.2\times 10^{-11}$ & 107.57 &  10.60 &  12.83 &  139.54 &  389.49 &  1546.79 &  10860.32 &  3167.75\\
    $3\times 10^{-11}$ & 89.19 &  29.62 &  8.68 &  102.43 &  253.44 &  1852.73 &  7044.59 &  1455.57\\
    $1.8\times 10^{-11}$ & 6.21 & -3.36 & -1.42 &  20.69 &  86.69 &  914.73 &  2792.53 &  572.07\\
    $1\times 10^{-11}$ & 1.15 & -1.32 & -1.67 &  12.19 &  33.08 &  392.81 &  1220.73 &  268.79\\
    $7.8\times 10^{-12}$ & 2.21 &  9.71 &  10.10 & -1.03 &  50.76 &  250.57 &  755.04 &  149.08\\
    $6\times 10^{-12}$ & -2.20 & -2.54 & -2.80 & -3.05 &  52.38 &  261.44 &  596.83 &  96.06\\
    \hline
  \end{tabular}
  \caption{$\Delta\mathrm{DICs}$ when comparing models P and 0.}
  \label{tab:0vsP}
\end{table}
\begin{table}[t]
  \centering
  \small
  \tabcolsep=0.09cm
  \begin{tabular}{|c| D{.}{.}{4.2} D{.}{.}{4.2} D{.}{.}{4.2} D{.}{.}{4.2} D{.}{.}{4.2} D{.}{.}{6.2} D{.}{.}{6.2} D{.}{.}{6.2} |}
    \hline
    & \multicolumn{8}{c|}{$f_\mathrm{pt}$ [Hz]} \\
    \hline
    $\Omega_\mathrm{pt} $ & \multicolumn{1}{c}{$4.2\times 10^{-4}$} & \multicolumn{1}{c}{$6\times 10^{-4}$} & \multicolumn{1}{c}{$7.8\times 10^{-4}$} & \multicolumn{1}{c}{$1\times 10^{-3}$} & \multicolumn{1}{c}{$1.8\times 10^{-3}$} & \multicolumn{1}{c}{$3\times 10^{-3}$} & \multicolumn{1}{c}{$6\times 10^{-3}$} & \multicolumn{1}{c|}{$1\times 10^{-2}$} \\
    \hline
    $3\times 10^{-10}$ & 3054.99 &  1623.88 &  696.48 &  716.57 &  2472.86 &  219023.16 &  688391.88 &  405296.24\\
    $1\times 10^{-10}$ & 813.02 &  499.69 &  130.18 &  195.17 &  1001.65 &  48885.84 &  134008.86 &  49604.82\\
    $6\times 10^{-11}$ & 381.35 &  316.77 &  79.84 &  86.54 &  958.92 &  21418.93 &  50343.05 &  10094.60\\
    $4.2\times 10^{-11}$ & 252.07 &  174.71 &  78.84 &  32.68 &  662.40 &  3158.71 &  24930.23 &  3154.20\\
    $3\times 10^{-11}$ & 135.92 &  101.03 &  46.32 &  5.16 &  686.02 &  3275.84 &  8040.96 &  1319.38\\
    $1.8\times 10^{-11}$ & 26.20 &  29.45 &  26.19 & -3.21 &  418.65 &  1998.05 &  3230.40 &  430.36\\
    $1\times 10^{-11}$ & 11.42 &  41.43 &  14.70 & -2.83 &  146.36 &  1033.88 &  1267.23 &  175.67\\
    $7.8\times 10^{-12}$ & 10.76 &  12.99 & -2.93 & -1.88 &  148.33 &  596.23 &  626.92 &  74.39\\
    $6\times 10^{-12}$ & 19.02 &  22.08 &  4.88 & -3.33 &  76.92 &  408.13 &  539.95 &  40.48\\
    \hline
  \end{tabular}
  \caption{$\Delta\mathrm{DICs}$ when comparing models Mc and MPc.}
  \label{tab:McvsMPc}
\end{table}

\bibliography{pt_gw}

\end{document}